\documentclass[11pt,letterpaper]{article}
\usepackage[T1]{fontenc}
\usepackage{lmodern}
\usepackage[margin=1in]{geometry}
\usepackage{placeins,graphicx,booktabs,mathtools,amssymb,amsthm,threeparttable,tcolorbox}
\usepackage[hyphens]{url}
\usepackage[round,authoryear]{natbib}
\usepackage[hidelinks,unicode]{hyperref}
\hypersetup{pdftitle={Bridging in Participatory Budgeting},
  pdfauthor={Tzeh Yuan Neoh, Jay Chooi, Anagha Satish, Alec Laprevotte}}
\newcommand{\metricfigurewidth}{.83\linewidth}
\newtheorem{theorem}{Theorem}[section]
\newtheorem{proposition}[theorem]{Proposition}
\newtheorem{lemma}[theorem]{Lemma}
\newtheorem{corollary}[theorem]{Corollary}
\theoremstyle{definition}
\newtheorem{definition}[theorem]{Definition}

\theoremstyle{remark}
\newtheorem{remark}[theorem]{Remark}

\title{Bridging in Participatory Budgeting}
\author{Tzeh Yuan Neoh%
        \thanks{Tzeh Yuan Neoh and Jay Chooi contributed equally as co-first authors.}\hspace{.45em}%
        \thanks{Correspondence to Tzeh Yuan Neoh at
          \href{mailto:tzehyuan_neoh@g.harvard.edu}{\texttt{tzehyuan\_neoh@g.harvard.edu}}.}\quad
        Jay Chooi\footnotemark[1]\quad
        Anagha Satish\quad Alec Laprevotte\\[0.45em]
        \normalsize Harvard University}
\date{}
\newcommand{\pairs}{\text{PAIRS}}

\newcommand{\cons}{\text{CONS}}
\newcommand{\av}{\text{AV}}
\newcommand{\cc}{\text{CC}}
\newcommand{\ejr}{\text{EJR}}

\newcommand{\cost}{\text{cost}}
\newcommand{\DP}{\textbf{DP}}
\newcommand{\OPT}{\text{OPT}}
\newcommand{\OurProblem}{\textnormal{\textsc{BiCri-Voting}}}
\newcommand{\clique}{\textnormal{\textsc{MinMax Clique}}}
\providecommand{\metricfigurewidth}{\linewidth}

\begin{document}

\maketitle

\begin{abstract}
Participatory budgeting funds projects of different costs from a shared budget.
We study two objectives for approval ballots. PAIRS counts voter pairs who approve a common funded project, and CONS also counts pairs linked by chains of shared approvals.
Both objectives admit polynomial-time algorithms on voter--candidate interval profiles. Given \(d\) projects whose deletion yields such a profile, PAIRS is fixed-parameter tractable in \(d\) and CONS is in XP. With only the deletion distance, both are in XP.
We ask whether a given election admits an outcome meeting prescribed bridging and representation guarantees. For either bridging objective, this problem is \(\Theta_2^P\)-complete for Approval Voting and Chamberlin--Courant and \(\Sigma_2^P\)-complete for Extended Justified Representation.
In 338 small-project Pabulib elections, a majority admit a joint optimum for each pairing with Approval Voting or Chamberlin--Courant. All 338 admit an outcome maximizing both CONS and Chamberlin--Courant, although standard rules often miss it.

\end{abstract}

\section{Introduction}
\label{intro}
Committee selection and participatory budgeting mechanisms are widely used to
allocate resources, choose representatives, and identify jointly supported
policies.  Yet, when a society is polarized, different groups may approve
almost disjoint sets of options, making it difficult to select an outcome that
connects rather than further separates them.

Political polarization has become an increasingly prominent global concern
\citep{polarizationMyth, levin2021dynamics, wilson2020polarization}.
Polarization is commonly understood as the division of a population into
groups with sharply divergent views \citep{fiorina2008political}.  Such
fragmentation can impede consensus and compromise, creating serious obstacles
to collective decision-making \citep{pnasnexus}.  This motivates the design of
mechanisms that not only represent diverse groups, but also create connections
between them.

Approval-based committee selection has traditionally focused on excellence,
diversity, or proportionality: selecting highly approved candidates
\citep{ElkindFSS15}, representing diverse opinions \citep{Chamberlin}, or
treating cohesive voter groups proportionally \citep{propwelfarism}.  These
criteria do not directly measure how a selected committee connects different
groups of voters.  A committee may represent voters well individually while
leaving them divided into disconnected groups.

\citet{dong_selecting_2025} address this gap through the notion of an
\emph{interlacing committee}.  They introduce two quantitative bridging
objectives: \pairs{}, which counts pairs of voters directly connected by a
jointly approved committee member, and \cons{}, which also counts pairs
connected through chains of shared approvals.  Their main theoretical
contributions are twofold.  First, they show that maximizing either objective
is NP-hard in general, but polynomial-time solvable under the one-dimensional
voter--candidate interval (VCI) domain.  Second, they construct instances that
exhibit severe worst-case approximation trade-offs between their bridging
objectives and other common objectives.

In their discussion of future work, \citet{dong_selecting_2025} explicitly
identify participatory budgeting as a natural setting in which to extend
their bridging objectives.  Participatory Budgeting (PB) is a real-world
mechanism in which communities allocate public funds among projects of varying
costs \citep{IndivisiblePB}.  Bridging has a particularly concrete
interpretation in this setting: funded projects can create shared spaces and
services through which residents from different groups interact.  Residents of
different regions or socioeconomic groups may support different projects, so
selecting projects that connect these groups may contribute to community
building alongside the usual goals of representation and proportionality.

Project costs change which connections a budget can cover.
In the unit-cost model, one can replace a candidate by another approved by
all of its supporters. In PB, only the first project may fit the budget. A
second question arises even when the project costs are known. The worst-case
trade-offs of \citet{dong_selecting_2025} leave open whether the two goals can
be attained together in the particular election facing a community.

\subsection{Our Contribution}
In Section~\ref{sec:PB_setting}, we show that either bridging objective can
be maximized in polynomial time on the voter--candidate interval (VCI)
domain. We give a direct dynamic program for \pairs{} and adapt the
frontier-component idea of \citet{dong_selecting_2025} for \cons{}. Both
programs minimize the cost of reaching each score. Their marginal-gain
bounds are exact along a minimum-cost solution. If a set of \(d\) projects
whose deletion yields VCI is supplied, we obtain an FPT algorithm for
\pairs{} and an XP algorithm for \cons{} in \(d\). Without a supplied set,
both objectives admit XP algorithms in the minimum deletion distance.

In Section~\ref{sec:complexity}, we classify the complexity of
\emph{instance-specific feasibility}. Given two guarantees, can any feasible
outcome meet both? For either bridging objective, the problem is
\(\Theta_2^P\)-complete when the representation objective is Approval Voting
(AV) or Chamberlin--Courant (CC), and \(\Sigma_2^P\)-complete for Extended
Justified Representation (EJR). The two hardness reductions use the budget
to encode, respectively, a comparison of optimum values and a choice that
must survive every coalition's EJR challenge.

In Section~\ref{sec:experiments}, we distinguish what is attainable from
what existing voting rules select in 338 Pabulib elections with at most
13 projects and 10{,}000 voters. For each of the four pairings with AV or CC,
a majority of elections admit a joint optimum. All 338 admit a joint
\cons{}--\cc{} optimum. We explain how the connectivity of the covered
voters relates these two scores and examine how close the rules come when
they miss an optimum.

\section{Related Work}

\paragraph{Approval Based Multiwinner Voting and Participatory Budgeting.}

Approval-based multiwinner voting has been extensively studied as a model for committee selection \citep{lackner2023multi}. Classical objectives include Approval Voting (AV) and the Chamberlin-Courant rule (CC). Further proportionality notions such as Extended Justified Representation (EJR) \citep{JustifiedRepresentationApproval} have been proposed to guarantee group-level fairness. The AV and CC objectives, along with a mild relaxation of EJR, which requires that EJR holds ``up to one project'', were studied for PB \citep{peters2021proportional}. 

\citet{brill2023pb} study how different notions of voter satisfaction affect proportionality in approval-based PB and when their requirements can be met together. \citet{serramia2024alignment} examine how to balance voter support with the alignment of funded projects to public values. Here the second objective measures connections among voters, and we ask whether prescribed guarantees can be met in a given election.

\paragraph{Selecting Interlacing Committees.}
Our work builds directly on the interlacing-committee framework of
\citet{dong_selecting_2025}, whose main results are summarized in the
introduction.  We move from unit-cost committee elections to participatory
budgeting with heterogeneous project costs, and show how VCI tractability can
be recovered despite this additional constraint.  We also study nearly VCI
instances, determine the complexity of instance-specific bicriteria
feasibility, and empirically examine whether the worst-case trade-offs arise in
real-world PB data.

\paragraph{Polarization and Bridging.}

Significant work has been done on the study of polarization in other social contexts. Social networks, built to entertain their users by connecting them with like-minded content, have been shown to increase polarization of users \citep{guerra2013measure,cinelli2021echo}. In this environment, coined a "filter bubble", small changes by a network administrator to minimize disagreement can greatly increase polarization \citep{chitra2020analyzing}.

The commonly proposed solution to these filter bubbles is analogous to our idea of bridging, where researchers make adjustments to the social network structure. Popular methods include adjustments to edge weights \citep{ninomiya2025mitigating} or the creation of new edges between disconnected components \citep{racz2023towards}. Other works introduce multiple objectives with the aim of reducing both polarization and disagreement on the network \citep{musco2018minimizing,kishore2024finding}. Our work further explores polarization mitigation with the use of bridging objectives in a new setting: participatory budgeting.

\section{Preliminaries}
\label{prelims}
We first introduce the notation used throughout the paper, and then define the bridging and representation axioms as well as the complexity classes $\Theta_2^P$ and $\Sigma_2^P$. For $k \in \mathbb{N}$, let $[k] = \{1,\dots,k\}$.

\subsection{Participatory Budgeting (PB)}

Let $P$ be a set of $m$ projects and $N=\{1,\dots,n\}$ a set of voters. Each project $p \in P$ has a cost $c(p) \in \mathbb{Q}_{>0}$, and there is a total budget limit $b \in \mathbb{Q}_{>0}$. Each voter $i \in N$ submits an approval ballot $A_i \subseteq P$, indicating the set of projects approved by voter $i$. For a project $p \in P$, let $N_p = \{\, i \in N \mid p \in A_i \,\}$ denote the set of voters approving $p$. For \(T\subseteq P\), write \(\cost(T)=\sum_{p\in T}c(p)\), and use \(\cost(p)=c(p)\) for a single project. An instance of approval-based participatory budgeting (PB) is given by the tuple $I = (P,A,b,c)$. A feasible \emph{outcome} for an instance $I$ is any subset $P' \subseteq P$ such that $\sum_{p \in P'} c(p) \le b$. When all projects have unit cost, this model coincides with approval-based multiwinner voting.

In addition to the general PB setting, we consider instances with spatial one-dimensional preferences. Following \cite{Godziszewski_Batko_Skowron_Faliszewski_2021}, an instance $I = (P,A,b,c)$ belongs to the voter--candidate interval (VCI) domain if there exist positions $\{x_p\}_{p \in P} \cup \{x_v\}_{v \in N} \subseteq \mathbb{R}$ and non-negative radii $\{r_p\}_{p \in P} \cup \{r_v\}_{v \in N} \subseteq \mathbb{R}_{\ge 0}$ such that, for all $v \in N$ and $p \in P$, it holds that $p \in A_v$ if and only if $|x_p - x_v| \le r_p + r_v$. Equivalently, voter $v$ approves project $p$ if and only if the intervals $[x_p - r_p, x_p + r_p]$ and $[x_v - r_v, x_v + r_v]$ intersect.

Write \(I_z=[x_z-r_z,x_z+r_z]\) for the \emph{approval interval} of voter or
project \(z\). Within a fixed VCI representation, project \(p\) \emph{dominates}
\(p'\) if \(I_{p'}\subsetneq I_p\). This geometric containment implies
\(N_{p'}\subseteq N_p\). An outcome is \emph{domination-free} if it contains
no such pair. A street or transit corridor illustrates the domain: a project
serves a stretch of the corridor and a resident approves projects that
intersect the stretch they can reach.

\subsection{Bridging and Representation Objectives}
Introduced by \cite{dong_selecting_2025}, the objectives \pairs{} and \cons{} quantify polarization by measuring how well an outcome connects voters. Intuitively, \pairs{} counts voter pairs that are directly connected through a single selected project, while \cons{} counts voter pairs that are connected indirectly via a sequence of selected projects. Formally, let \(N^{(2)} = \{\{u,v\} \subseteq N \mid u \neq v\}\) denote the set of unordered pairs of voters. For an outcome \(P' \subseteq P\), the two objectives are defined as:
\begin{itemize}
    \item \(\pairs(P') = \bigl|\{\{i,j\} \in N^{(2)} \mid A_i \cap A_j \cap P' \neq \emptyset\}\bigr|\).
    \item \(\cons(P') = \bigl|\{\{u,v\} \in N^{(2)} \mid u \sim_{P'} v\}\bigr|\), where \(u \sim_{P'} v\) if and only if there exists a sequence of voters \(v_1,\ldots,v_\ell \in N\) with \(u = v_1\), \(v = v_\ell\), and
    \[
        A_{v_i} \cap A_{v_{i+1}} \cap P' \neq \emptyset
        \quad \text{for all } i \in \{1,\ldots,\ell-1\}.
    \]
\end{itemize}
In addition to these bridging notions, we consider standard approval-based representation objectives capturing \emph{excellence}, \emph{diversity}, and \emph{proportionality} \citep{faliszewski2017multiwinner}.

Both excellence and diversity are defined quantitatively. The \emph{approval score} (\av) measures the total number of approvals received by the projects in an outcome and is defined as
\[
\av(P') := \sum_{v \in N} |A_v \cap P'|.
\]
The \emph{Chamberlin--Courant score} (\cc) measures the number of voters who approve at least one selected project and is defined as
\[
\cc(P') := \bigl|\{\, v \in N \mid A_v \cap P' \neq \emptyset \,\}\bigr|.
\]

For any score \(f\in\{\pairs,\cons,\av,\cc\}\), let
\(\OPT_f=\max_{\cost(W)\le b}f(W)\). An outcome \(W\) achieves
\(\alpha\)-\(f\) if \(f(W)\ge\alpha\OPT_f\). We write
\(\alpha_f(W)=f(W)/\OPT_f\) for its attained ratio, taking it to be one
when \(\OPT_f=0\).

Proportionality is captured by the axiom of \emph{Extended Justified Representation} (\ejr) introduced by \cite{JustifiedRepresentationApproval}. For $\alpha \in [0,1]$, an outcome $P'$ satisfies $\alpha$-EJR if, for every nonempty group of voters $S \subseteq N$ and every set of projects $T \subseteq P$ such that all voters in $S$ approve all projects in $T$ (i.e., $T \subseteq \bigcap_{i \in S} A_i$) and $T$ can be afforded by $S$ using its $\alpha$-scaled proportional share of the budget (i.e., $\operatorname{cost}(T) \le \alpha \cdot \frac{|S|}{|N|} \cdot b$), there exists at least one voter $i \in S$ such that
\[
\operatorname{cost}(A_i \cap P') \ge \operatorname{cost}(T).
\]

Ordinary EJR is \(1\)-EJR. At \(\alpha=0\), every outcome satisfies the condition because projects have positive costs.

\subsection{Complexity Classes}
For input length \(L\) and integer parameter \(k\), fixed-parameter
tractability (FPT) means a running time \(g(k)L^{O(1)}\). The class XP
permits a running time \(g(k)L^{h(k)}\), for computable functions \(g,h\).
The class \(\Theta_2^P=\mathrm{P}^{\mathrm{NP}}_{\parallel}\) permits a
polynomial-time algorithm a single round of parallel NP queries, or
equivalently a logarithmic number of adaptive queries in the input length
\citep{wagner1990bounded}. The class
\(\Sigma_2^P=\mathrm{NP}^{\mathrm{NP}}\) permits a nondeterministic
polynomial-time algorithm access to an NP oracle
\citep{papadimitriou1994complexity}. Section~\ref{sec:complexity}
relates these two classes to the structure of the representation guarantee.

\section{Computing Bridging Objectives in Participatory Budgeting}\label{sec:PB_setting}

We first consider the decision problems associated with maximizing PAIRS and
CONS.
\begin{tcolorbox}[title=\textsc{MAXIMIZING PAIRS / MAXIMIZING CONS (PAIRS-DEC / CONS-DEC)}]

\textbf{Input}: An instance $I = (P,A,b,c)$ and a target value $\kappa$.

\tcblower
\textbf{Question}: Does there exist a feasible outcome $P' \subseteq P$ with
$\sum_{p \in P'} c(p) \leq b$ such that
\[
    \textsc{PAIRS}(P') \ge \kappa
    \quad\text{(for PAIRS-DEC)},
\]
or
\[
    \textsc{CONS}(P') \ge \kappa
    \quad\text{(for CONS-DEC)}?
\]

\end{tcolorbox}

For both PAIRS and CONS, computing a committee that maximizes the objective is
NP-hard \citep{dong_selecting_2025}.  Since approval-based multiwinner voting is
the unit-cost special case of participatory budgeting, this hardness carries
over immediately to PB.

\begin{proposition}
   Both PAIRS-DEC and CONS-DEC remain NP-complete for general PB instances.
\end{proposition}

\paragraph{From committees to budgets.}
On a unit-cost VCI profile, \citet{dong_selecting_2025} first remove candidates
whose supporters all approve another candidate, yielding a candidate-interval
profile. They solve \pairs{} by applying a Chamberlin--Courant algorithm to
the associated voter-pair instance. For \cons{}, their dynamic program tracks
committee size and the size of the component containing the last selected
candidate. With heterogeneous costs, deleting a dominated project up front can
remove the only affordable option. Indexing a dynamic program by the
integer-scaled budget can also require pseudo-polynomial time.

Our \pairs{} program works directly on the voter and project intervals.
For \cons{} we retain the frontier component and keep a lower bound on its
size. Both programs record the minimum cost of reaching each score, of which
there are at most \(1+\binom n2\). If a selected project dominates another
selected project, the latter can be dropped without losing any connection.
Thus a minimum-cost outcome for a score threshold is domination-free. We show
that the marginal-gain lower bounds used by the programs are exact on such
outcomes.

\paragraph{Obtaining the interval representation.}
The bipartite graph with vertex classes \(N\) and \(P\) has edge \(vp\) when
\(p\in A_v\). It is an interval bigraph exactly when the profile is VCI:
intervals from opposite classes intersect precisely on the approval edges
\citep{avramidis2026interval}. \citet{muller1997interval} construct an
interval representation in polynomial time whenever one exists. Its midpoints
and endpoints supply the orders for both algorithms from the ordinary PB input.

\begin{theorem}\label{thm:PAIRS_poly}
    In the VCI domain, there is a polynomial-time algorithm for PAIRS-DEC.
\end{theorem}
\begin{proof}[Proof sketch]
Discard projects with fewer than two supporters. They add no voter pair or
connection, so every bridging optimum remains attainable.
Arrange the projects as $p_1,\dots,p_m$ such that $i \le j$ implies
$x_{p_i} \le x_{p_j}$, and arrange the voters as $v_1,\dots,v_n$ such that
$i \le j$ implies $x_{v_i}-r_{v_i} \le x_{v_j}-r_{v_j}$.

\paragraph{Calculating the gain in \pairs\ from adding a project.}
Let $P'$ be an outcome whose maximum project index is $j$
(i.e., $P' \subseteq \{p_1,\dots,p_j\}$ and $p_j \in P'$), and let $i$ be the
maximum index of any voter covered by $P'$.
Consider adding a project $p_k$ with $k>j$.

Let $N_{p_k}$ denote the set of voters approving $p_k$, and partition it into
$N_{p_k}^{\le i} := N_{p_k} \cap \{v_1,\dots,v_i\}$ and
$N_{p_k}^{> i} := N_{p_k} \cap \{v_{i+1},\dots,v_n\}$.
Define the marginal gain
$\pairs(p_k \mid P') := \pairs(P' \cup \{p_k\}) - \pairs(P')$.

We first derive a lower bound on $\pairs(p_k \mid P')$ and then show that this
bound is tight whenever $P' \cup \{p_k\}$ contains no dominated project.

\begin{lemma}
$\pairs(p_k \mid P') \ge \binom{|N_{p_k}|}{2} - \binom{|N_{p_k}^{\le i}|}{2}$.
Moreover, if $P' \cup \{p_k\}$ contains no dominated project, then
\(
\pairs(p_k \mid P')
=
\binom{|N_{p_k}|}{2} - \binom{|N_{p_k}^{\le i}|}{2}.
\)
\end{lemma}

\paragraph{Defining the dynamic program.}
For $j \in \{0,\dots,m\}$, $i \in \{0,\dots,n\}$, and
$q \in \{0,\dots,\binom{n}{2}\}$, let $\mathbf{min}(i,j,q)$ denote the minimum
cost of an outcome $P' \subseteq \{p_1,\dots,p_j\}$ such that:
(i) $\pairs(P') \geq q$, and
(ii) the maximum index of any covered voter is exactly $i$
(that is, $v_i$ is covered and no voter in $\{v_{i+1},\dots,v_n\}$ is covered).
If no such outcome exists, set $\mathbf{min}(i,j,q)=\infty$.

We define a DP table $\DP(i,j,q)$ intended to equal $\mathbf{min}(i,j,q)$.
Initialize $\DP(0,0,0)=0$ and $\DP(i,j,q)=\infty$ for all other states.

We fill the table for $j=1,\dots,m$.
For each state $(i,j,q)$, skipping project $p_j$ yields
\[
\DP(i,j,q) := \min\{\DP(i,j,q),\, \DP(i,j-1,q)\}.
\]
Now consider including $p_j$.
Let $\rho_j := \max\{\ell \in \{1,\dots,n\} : v_\ell \in N_{p_j}\}$ and
$i^+ := \max\{i,\rho_j\}$.
For each $q$ with $\DP(i,j-1,q) \ne \infty$, define
\[
\Delta_j(i):=
\binom{|N_{p_j}|}{2} - \binom{|N_{p_j}^{\le i}|}{2}.
\]
For every $\ell\in\{0,\dots,\Delta_j(i)\}$ with
$q+\ell\le \binom{n}{2}$, we update
\begin{align*}
\DP(i^+,j,q+\ell) :=\;&
\min\{\DP(i^+,j,q+\ell),\\
&\DP(i,j-1,q) + \cost(p_j)\}.
\end{align*}

Every update credits only pairs that the project connects. Conversely, along
a minimum-cost witness for a threshold, the lemma makes every credited gain
exact. Induction on the project prefix therefore gives the desired minimum
cost. The answer is yes exactly when \(\min_i\DP(i,m,\kappa)\le b\)
(for \(\kappa>\binom{n}{2}\), the answer is no).
\end{proof}

For \cons{}, the number of new pairs also depends on the size of the component
that a new project joins. This is the frontier information used by
\citet{dong_selecting_2025}.

\begin{theorem}\label{thm:CONS_poly}
    In the VCI domain, there is a polynomial-time algorithm for CONS-DEC.
\end{theorem}
\begin{proof}[Proof sketch]
Order the projects and voters as in the proof of Theorem~\ref{thm:PAIRS_poly}.
Let $P'$ be an outcome whose maximum project index is $j$, and let $i$ be the
maximum index of any voter covered by $P'$. The voters covered by $P'$ are
partitioned into equivalence classes under the connectivity relation induced by
$P'$; let $E_i$ denote the class containing voter $v_i$.

Consider adding a project $p_k$ with $k>j$. If $P'\cup\{p_k\}$ contains no
dominated project, then the marginal contribution of $p_k$ to $\cons$ depends
only on whether $p_k$ connects to an already-covered voter and, if so, on the
size of the frontier class $E_i$: if $p_k$ connects to no covered voter, it
creates a new equivalence class and contributes $\binom{|N_{p_k}|}{2}$; otherwise,
all voters approving $p_k$ merge into $E_i$, and the contribution is
$\binom{|N_{p_k}^{>i}|}{2}+|N_{p_k}^{>i}|\cdot |E_i|$.

This characterization allows a dynamic program similar to that for
Theorem~\ref{thm:PAIRS_poly}, augmented to track a lower bound on the size of
the frontier equivalence class.  The state is
$\DP(i,j,q,w)$, where $i$ is the maximum covered voter index, $q$ is the
\cons{} threshold, and $w$ is the stored lower bound on the size of $E_i$.
If $p_j$ creates a new component, then $w$ is updated to $|N_{p_j}|$; if it
extends the frontier component, then the marginal contribution is
$\binom{|N_{p_j}^{>i}|}{2}+|N_{p_j}^{>i}|\cdot w$ and the new frontier-size
lower bound is $w+|N_{p_j}^{>i}|$.
Correctness follows by induction on the project index, using the above
characterization of marginal gains.
\end{proof}

\subsection{Nearly One-Dimensional Preferences}

A few projects may draw support from distant parts of an otherwise ordered
electorate. We measure the departure from VCI by the number of projects whose
deletion restores an interval representation.

\begin{definition}[Candidate Deletion Distance]
Let $I=(P,A,b,c)$ be an instance.  For $D\subseteq P$, let
\[
    I-D
    :=
    \bigl(P\setminus D,\,(A_v\setminus D)_{v\in N},\,b,\,
    c|_{P\setminus D}\bigr)
\]
be the instance obtained by deleting the projects in $D$.  The candidate
deletion distance of $I$ to the VCI domain is
\[
    \delta_{\mathrm{VCI}}(I)
    =
    \min \bigl\{|D| : D \subseteq P,\; I-D
    \text{ is a VCI instance}\bigr\}.
\]
Thus, $\delta_{\mathrm{VCI}}(I)$ is the smallest number of projects that must
be removed from $I$ so that the remaining instance satisfies the VCI property.
\end{definition}

For comparison, consider the total costs of making ballots intervals on a
common project order by flips, ballot completion, adjacent swaps, or the
number of forbidden triples \citep{delemazure2024comparing}. Minimizing over
orders gives each of the four distance measures below.

\begin{proposition}\label{prop:distance_comparison}
Candidate deletion distance to VCI is at most each of minimum flips, ballot
completion, forbidden triples, and minimum swaps. It is incomparable with
voter deletion distance to VCI.
\end{proposition}
\begin{corollary}\label{cor:distance_transfer}
Given a project order and a repair of cost \(k\) for any of the four measures
in Proposition~\ref{prop:distance_comparison}, one can construct a VCI
deletion set of size at most \(k\) in polynomial time. Thus the algorithms
below give FPT for \pairs{} and XP for \cons{} in the supplied repair cost.
With only a bound on that cost, both problems are in XP.
\end{corollary}
Appendix~\ref{app:proof-distance-comparison} gives the construction. An upper
bound on the distance alone does not supply the deletion set required for
the FPT running time.

\begin{theorem}\label{thm:FPT_PAIRS}
Given a set \(D\subseteq P\) such that \(I-D\) is VCI, PAIRS-DEC can be
solved in \(2^{|D|}\operatorname{poly}(|I|)\) time.
\end{theorem}
\begin{proof}[Proof sketch]
Enumerate the subsets \(D'\subseteq D\) with \(\cost(D')\le b\). In this
branch, select exactly \(D'\) from \(D\) and use the remaining budget
\(b-\cost(D')\) on the VCI profile \(I-D\).

We then optimize a modified objective $\pairs_{-D'}$, which counts only voter
pairs not already covered by $D'$.
Let
\[
\mathit{cov}_{D'} =
\{\{i,j\} \in N^{(2)} : A_i \cap A_j \cap D' \neq \emptyset\}
\]
be the set of voter pairs already covered by $D'$.
For \(W\subseteq P\setminus D\), let \(\mathit{cov}_W\) denote its
analogous set of covered pairs. Define
\[
  \pairs_{-D'}(W)=|\mathit{cov}_W\setminus\mathit{cov}_{D'}|.
\]

For the branch corresponding to $D'$, maximizing $\pairs$ in the original
instance is equivalent to maximizing $\pairs_{-D'}$ over the remaining VCI projects.  Thus we can use
the DP of Theorem~\ref{thm:PAIRS_poly}, changing only the marginal-gain
calculation so that it ignores pairs already covered by $D'$.  If \(W\) maximizes the modified objective within the remaining budget,
the branch succeeds exactly when
\[
   |\mathit{cov}_{D'}|+\pairs_{-D'}(W)\ge\kappa.
\]
It suffices that one branch succeeds.
\end{proof}

The case of $\cons$ is more delicate.  Once projects in $D'$ are fixed, the
equivalence classes they induce can connect distant parts of the VCI order:
if voters $u$ and $v$ are connected through $D'$, then any VCI projects approved
by $u$ and by $v$ may become part of the same connected component.  Thus, unlike
in the pure VCI case, it is not enough to track a single frontier class.  We
track how the components induced by $D'$ merge, the sizes of the resulting
classes, and the component containing the current VCI frontier. With a supplied
set of size \(d\), this gives \(|I|^{O(d+1)}\) time up to a factor depending
only on \(d\).
\begin{theorem}\label{thm:XP_CONS}
    CONS-DEC is XP with respect to $\delta_{\mathrm{VCI}}(I)$.
\end{theorem}

\begin{remark}
Without a supplied set, enumerate candidate deletion sets in order of size
and test the residual approval graph with \citeauthor{muller1997interval}'s
recognition algorithm. If the minimum distance is \(k\), a witness is found
after testing at most \(\sum_{i=0}^k\binom mi\) sets. Combining this search
with the two algorithms gives XP for both objectives in \(k\), including
Theorem~\ref{thm:XP_CONS} on the ordinary PB input. Whether \cons{} is FPT
in the deletion distance remains open.
\end{remark}

\section{Feasibility of Simultaneously Achieving Bridging and Representation}
\label{sec:complexity}
\citet{dong_selecting_2025} establish worst-case trade-offs between bridging and representation even in the unit-cost setting. In particular, for $\pairs$ and any representation objective $\mathit{rep} \in \{\av,\cc,\ejr\}$, any voting rule that guarantees $\alpha$-$\pairs$ and $\beta$-$\mathit{rep}$ must satisfy $\alpha + \beta \leq 1$. For $\cons$, the trade-off is even stronger: any voting rule that guarantees $\alpha^2$-$\cons$ and $\beta$-$\mathit{rep}$ must satisfy $\alpha + \beta \leq 1$.

A community faces one election at a time. For its particular ballots,
costs and budget, are prescribed levels of bridging and representation
simultaneously attainable? We ask this feasibility question even when no
rule can guarantee the same levels on every election.

\begin{tcolorbox}[title={Bicriteria feasibility (\OurProblem)}]
\textbf{Input.} A PB instance \(I\), a bridging objective
\(BR\in\{\pairs,\cons\}\), a representation objective
\(REP\in\{\av,\cc,\ejr\}\), and rational thresholds \(\alpha,\beta\in[0,1]\).
\tcblower
\textbf{Question.} Is there a feasible outcome \(W\) that achieves both
\(\alpha\)-\(BR\) and \(\beta\)-\(REP\)?
\end{tcolorbox}

\begin{theorem}\label{thm:theta_2^P}
     For $BR \in \{\pairs, \cons\}$ and $REP \in \{AV, CC\}$, \OurProblem\, is $\Theta_2^P$-complete.
\end{theorem}
\begin{theorem}\label{thm:sigma_2^P}
     For $BR \in \{\pairs, \cons\}$ and $REP \in \{EJR\}$, \OurProblem\, is $\Sigma_2^P$-complete.
\end{theorem}
\paragraph{Comparing two optimum values.}
For AV and CC, both attained scores are computable in polynomial time.
The bridging optimum is at most \(\binom n2\), the AV optimum at most \(nm\),
and the CC optimum at most \(n\). Binary search with an NP oracle finds both
optima in \(O(\log(nm+1))\) queries. One further query tests whether a
single outcome meets both thresholds.

For hardness, the reduction first encodes \textsc{Compare-MaxSAT} as two
maximum-coverage instances whose optimum values must be compared. The
unconstrained bridging optimum in the PB instance is determined by expensive
projects encoding the second coverage instance. Meeting the representation
threshold forces expenditure on projects with single supporters, leaving
enough budget only for unit-cost projects encoding the first instance.
Consequently the bridging threshold can be met along with the representation
threshold exactly when the first coverage optimum is at least the second.
Appendix~\ref{app:bicriteria-proofs} gives the constructions for both
bridging objectives.

\paragraph{Checking a coalition.}
For EJR, guess a feasible outcome. Failure of a positive bridging threshold
is certified by a better feasible outcome, and failure of \(\beta\)-EJR is
certified by a coalition and a commonly approved bundle. An NP oracle can
check for either certificate. For hardness, reduce from \textsc{Min--Max
Clique Deletion}. Bridging optimality forces a fixed core of projects and
exactly \(\ell\) vertex projects. Choosing those vertices encodes the
deletion decision. An EJR violation then exists exactly when the remaining
graph contains a clique of size \(k\). This gives hardness already for
\(\alpha=\beta=1\). A related existential choice followed by a universal
test appears in allocation with envy-freeness and Pareto efficiency
\citep{de2009complexity}.

\paragraph{Solver implications.}
For AV or CC, optimize each score separately, then impose both resulting
thresholds in one mixed-integer feasibility model. For EJR, a single
polynomial-size ILP constructible in polynomial time for the joint decision
problem would collapse the polynomial hierarchy. For a fixed tentative
outcome, however, a violating coalition can be sought by an integer program.
One can add a constraint for each witnessed coalition and bundle and repeat
until a candidate passes or the model is infeasible. Appendix~\ref{sec:ejr-test}
gives the separation test. The number of constraints generated can be
exponential.

\begin{remark}
Ratios compare outcomes against what their own ballots and budget permit.
With absolute thresholds for two quantitative scores, the decision problem
is NP-complete: an outcome certifies feasibility, and setting the
representation threshold to zero recovers the single-objective bridging
problem. For EJR, the construction in Lemma~\ref{lem:sigma_2^P} also yields
hardness with an absolute bridging threshold: evaluating the fixed core
together with any \(\ell\) vertex projects gives the optimum in polynomial
time. Both reductions use heterogeneous costs. The corresponding bicriteria
classifications for unit costs remain open.
\end{remark}

\section{Bridging in Real-World Participatory Budgeting Instances}
\label{sec:experiments}

\paragraph{Elections and comparisons.}
We study 338 of the 835 approval-based elections in the Pabulib snapshot
\citep{Faliszewski2023ParticipatoryBudgeting}. We retain nonexperimental
elections with at most 13 projects and 10{,}000 voters.\footnote{Pabulib
snapshot \url{https://pabulib.org/download/snapshot/98659fc686f297cc}.}
The cutoff reflects our implementation: there are \(2^m\) candidate subsets
before pruning by cost, and scoring one feasible outcome for PAIRS can require
an \(n\times n\) matrix. One further project doubles the number of candidate
subsets. Enumeration lets us distinguish a genuine incompatibility in an
election from a rule's failure to find a jointly optimal outcome.

We compute AV, CC, PAIRS and CONS for every feasible outcome. For each
pairing with AV or CC, the \emph{attainable}
benchmark records whether the two sets of maximizers intersect, including
all ties. We compare this benchmark with the Method of Equal Shares (MES)
\citep{peters2021proportional} and five greedy rules: CC, PAV, AV, AV per
unit cost, and AV per squared cost. Appendix~\ref{subsec:implementation}
describes computation and tie handling. We also report the archived
cost-threshold screen for EJR as a diagnostic. The screen can flag a
shortfall when no indivisible bundle witnesses one. The empirical EJR
columns therefore do not certify the theoretical EJR axiom
(Appendix~\ref{sec:ejr-test}).

\subsection{Results}
\begin{figure}[tbp]
  \centering
  \includegraphics[width=\metricfigurewidth]{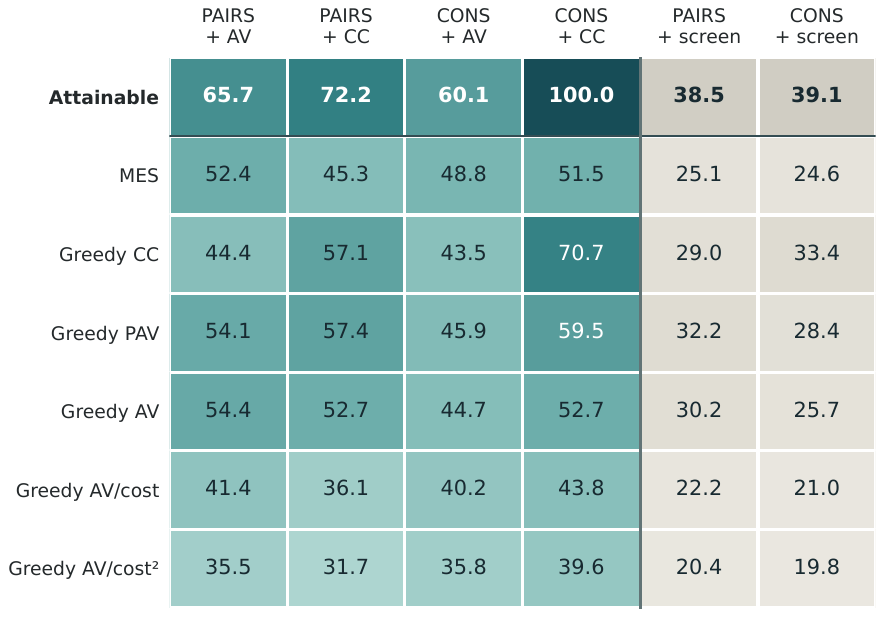}
  \caption{Percentage of the 338 elections in which the indicated targets
  are jointly reached. The first row allows any feasible outcome. Each other
  row uses the outcome selected by that rule. The last two columns use the
  EJR screen.}
  \label{fig:main_tradeoff_bar}
\end{figure}

A joint optimum exists for PAIRS and AV in 222 elections, PAIRS and CC in
244, CONS and AV in 203, and CONS and CC in all 338
(Figure~\ref{fig:main_tradeoff_bar}). These figures describe the retained
elections. They do not estimate compatibility in elections with more
projects or voters.

\paragraph{Why CONS and CC align.}
Write \(c=\cc(W)\) and \(C^*=\OPT_{\cc}\). Any pair counted by CONS consists
of two covered voters, so \(\cons(W)\le\binom c2\). If a voter approves every
project in \(W\), every covered voter is connected through that voter.
Hence \(\cons(W)=\binom c2\), and any such CC maximizer also maximizes CONS.
In 203 elections a CC maximizer has a common approver. More generally,
337 elections attain \(\OPT_{\cons}=\binom{C^*}{2}\); the remaining election
still has a joint optimum. When both \(W\) and a CC maximizer connect all
their covered voters and \(C^*\ge2\),
\[
 \alpha_{\cons}(W)=\frac{c(c-1)}{C^*(C^*-1)}
 \ \approx\ \alpha_{\cc}(W)^2.
\]
The median election has a common approver for 97.2\% of its nonempty
feasible outcomes. Appendix~\ref{subsec:additional-empirical-results}
shows the distribution across elections.

\begin{figure}[tbp]
  \centering
  \includegraphics[width=\metricfigurewidth]{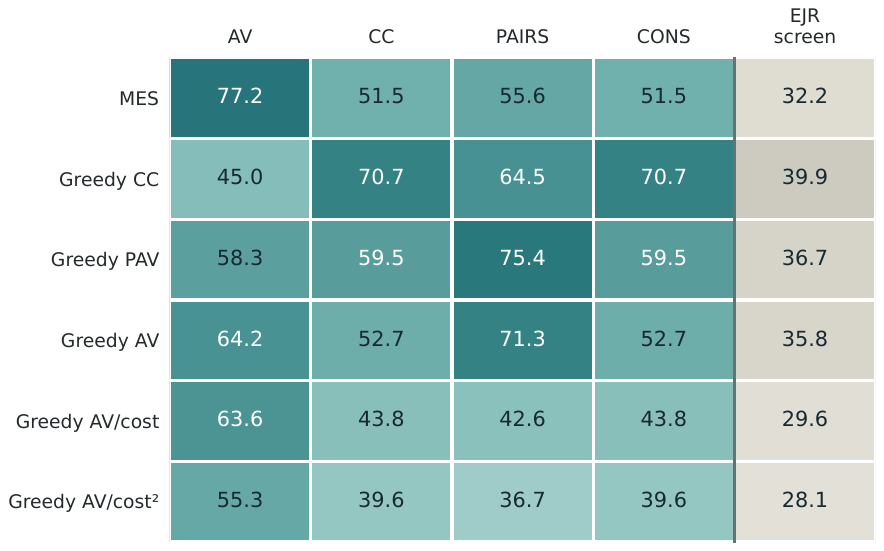}
  \caption{Percentage of elections in which a rule attains each
  single-score optimum or passes the EJR screen. Each column uses all
  338 elections.}
  \label{fig:metrics_bar}
\end{figure}

\paragraph{What the rules select.}
For example, greedy CC reaches a joint CONS--CC optimum in 239 of the
338 elections where one exists. An exact miss can still be close: among
the 99 elections where greedy CC misses the CONS optimum, its median CONS
ratio is 0.946. Equivalent comparisons for the other rules are in
Figures~\ref{fig:metrics_bar} and~\ref{fig:alpha_cdf}. MES targets
proportionality, and a greedy rule need not globally maximize its own
objective. None of these implementations uses bridging to resolve score ties.

The other objectives reward different changes to an outcome. Another project
for the same supporters can increase AV without adding any direct pair.
Joining two components of sizes \(a\) and \(b\) adds \(ab\) pairs to CONS
even if CC is unchanged. EJR protects cohesive groups according to their
budget shares. Neither bridging score tracks these entitlements. The
comparisons identify attainable outcomes and rule performance within this
sample; they do not isolate which election features caused a particular
rule to succeed.

\section{Conclusion}
We give algorithms for maximizing PAIRS and CONS on VCI profiles and on
profiles close to this domain. For general PB instances, the complexity of
joint feasibility depends on the representation requirement. In the 338
retained Pabulib elections, a majority admit a joint optimum for each
pairing with AV or CC. All admit a joint CONS--CC optimum. For EJR, an
exact empirical classification requires testing indivisible bundle costs.
Bridging-aware voting rules and elections with more projects are natural
next steps.

More discriminating objectives could distinguish multiple independent connections.  If an
instance contains a project approved by every voter, for example, every outcome
containing that project maximizes PAIRS and CONS, even though these outcomes may
differ substantially in how well their remaining projects connect otherwise
polarized groups.  The objectives therefore recognize the universal bridge but
not any additional, independent connections.  Refined objectives could account
for the number or strength of such connections while retaining the intuitive
interpretation of PAIRS and CONS.

Cost-aware notions of bridging also require care.  Consider three voter groups of sizes
\(0.4n\), \(0.3n\), and \(0.3n\).  A single project costing the entire budget
may connect only the first two groups, whereas three projects costing one third
of the budget each may collectively connect every pair of groups.  Nevertheless,
weighting each connected voter pair by the most expensive project or total cost of projects connecting the voter pair would favor the single expensive project.   Thus,
incorporating project costs naively may distort, rather than improve, the
notion of bridging; we develop these examples in
Appendix~\ref{app:cost-aware-bridging}.

\FloatBarrier
\bibliographystyle{plainnat}
\bibliography{references}
\clearpage
\appendix
\setcounter{secnumdepth}{2}
\section{Omitted Proofs for Bridging Objectives in Participatory Budgeting}
\subsection{Proof of Proposition \ref{prop:distance_comparison}}
\label{app:proof-distance-comparison}
\begin{proof}
Fix a common project order. For flips or ballot completion, delete each
project whose column contains an edited entry. The original and repaired
ballots then agree on the remaining order, on which each ballot is an
interval. For forbidden triples, delete the middle project of each occurrence
of the pattern approved--unapproved--approved. A triple in the restricted
order would already have been present before deletion. For adjacent swaps,
each ballot has a sequence of swaps taking the common order to an order in
which that ballot is an interval. Mark one of the two projects in each swap
and delete all marked projects. For every voter, the two orders now restrict
to the same order. In each construction we delete at most one project per
unit of repair cost. Represent the remaining projects as points and the
nonempty ballots as intervals to obtain VCI. Empty ballots can be placed
away from all project points.

For incomparability, use three projects \(a,b,c\) and the three ballots
\(\{a,b\},\{b,c\},\{a,c\}\). This profile is not VCI. Its project supporter
sets are pairwise incomparable, so no project interval can contain another.
Their left and right endpoints must therefore occur in the same order, and
every voter must approve a consecutive block in that order. No order makes
all three pairs consecutive. Deleting one voter or one project suffices;
on the order \(a,b,c\), swapping \(b,c\) for the voter approving \(\{a,c\}\)
repairs the only noninterval ballot. Thus the swap bound can be tight.
If each voter is replaced by \(t\) identical copies, one project deletion
still suffices, but at least \(t\) voter deletions are necessary to remove
all copies of one ballot type, and \(t\) suffice. Duplicating each project
instead gives the reverse distances. Duplicates can inherit the interval
of the object they copy.
\end{proof}

\subsection{Proof of Theorem \ref{thm:PAIRS_poly}}
Arrange the projects as $p_1,\dots,p_m$ such that $a \le b$ implies
$x_{p_a} \le x_{p_b}$.  Arrange the voters as $v_1,\dots,v_n$ such that
$a \le b$ implies $x_{v_a}-r_{v_a} \le x_{v_b}-r_{v_b}$.  For any voter or
project $z$, write
\[
L_z:=x_z-r_z
\qquad\text{and}\qquad
R_z:=x_z+r_z
\]
for the left and right endpoints of its interval.  We assume, without loss of
generality, that all projects under consideration have at least two supporters;
projects with at most one supporter never improve either \pairs{} or \cons{} and
can be deleted.

\paragraph{Calculating the gain in \pairs\ from adding a project.}
Let $P'$ be an outcome whose maximum project index is $j$ (i.e., $P' \subseteq \{p_1,\dots,p_j\}$ and $p_j \in P'$). Let $i$ be the maximum voter index among voters covered by $P'$ (equivalently, the largest index of any voter that approves at least one project in $P'$). Consider adding a project $p_k$ with $k>j$.

Let $N_{p_k}$ denote the set of voters approving $p_k$. Partition $N_{p_k}$ into $N_{p_k}^{\le i} := N_{p_k} \cap \{v_1,\dots,v_i\}$ and $N_{p_k}^{> i} := N_{p_k} \cap \{v_{i+1},\dots,v_n\}$. Define the marginal gain $ \pairs(p_k \mid P') := \pairs(P' \cup \{p_k\}) - \pairs(P')$.

We first give a lower bound on $\pairs(p_k \mid P')$.

\begin{lemma}\label{lem:PAIRS0}
$\pairs(p_k \mid P') \ge \binom{|N_{p_k}|}{2} - \binom{|N_{p_k}^{\le i}|}{2}$.
\end{lemma}
\begin{proof}
By the choice of $i$, every voter in $N_{p_k}^{> i}$ approves none of the projects in $P'$. Hence, before adding $p_k$, no pair consisting of one voter from $N_{p_k}^{\le i}$ and one voter from $N_{p_k}^{> i}$ is covered by $P'$, contributing $|N_{p_k}^{\le i}|\cdot |N_{p_k}^{> i}|$ uncovered pairs. Likewise, no pair of voters both in $N_{p_k}^{> i}$ is covered by $P'$, contributing $\binom{|N_{p_k}^{> i}|}{2}$ uncovered pairs. After adding $p_k$, all these pairs become covered. Therefore,
\begin{align*}
    \pairs(p_k \mid P') &\ge |N_{p_k}^{\le i}|\cdot |N_{p_k}^{> i}| + \binom{|N_{p_k}^{> i}|}{2}\\
&= \binom{|N_{p_k}|}{2} - \binom{|N_{p_k}^{\le i}|}{2}
\end{align*}

\end{proof}

\begin{lemma}\label{lem:PAIRS1}
If $P' \cup \{p_k\}$ contains no dominated project, then every voter $v \in N_{p_k}^{\le i}$ approves $p_j$.
\end{lemma}
\begin{proof}
Let $p_t\in P'$ be a project approved by the frontier voter $v_i$.  Suppose,
for contradiction, that some voter $v\in N_{p_k}^{\le i}$ does not approve
$p_j$.  Since $v$ does not approve $p_j$, the intervals of $v$ and $p_j$ are
disjoint.  Thus either $R_{p_j}<L_v$ or $L_{p_j}>R_v$.

First suppose that $R_{p_j}<L_v$.  Since $t\le j$ and $p_j$ is not dominated
by $p_t$ in $P'\cup\{p_k\}$, we have $R_{p_t}\le R_{p_j}$; otherwise the
interval of $p_t$ would strictly contain the interval of $p_j$.  Because
$v\le i$, we have $L_v\le L_{v_i}$.  Hence
$R_{p_t}\le R_{p_j}<L_v\le L_{v_i}$, contradicting that $v_i$ approves $p_t$.

It remains to consider $L_{p_j}>R_v$.  Since $j<k$ and $p_j$ is not dominated
by $p_k$, we have $L_{p_k}\ge L_{p_j}$; otherwise the interval of $p_k$ would
strictly contain the interval of $p_j$.  Hence
$L_{p_k}\ge L_{p_j}>R_v$, contradicting that $v$ approves $p_k$.
\end{proof}

\begin{lemma}\label{lem:PAIRS2}
If $P' \cup \{p_k\}$ contains no dominated project, then $\pairs(p_k \mid P') = \binom{|N_{p_k}|}{2} - \binom{|N_{p_k}^{\le i}|}{2}$.
\end{lemma}
\begin{proof}
Lemma~\ref{lem:PAIRS0} gives the lower bound. For the upper bound, by Lemma~\ref{lem:PAIRS1}, every voter in $N_{p_k}^{\le i}$ approves $p_j$, and hence any pair of voters both in $N_{p_k}^{\le i}$ is already covered by $P'$. Therefore, the only newly covered pairs after adding $p_k$ are those with at least one endpoint in $N_{p_k}^{> i}$, whose number is exactly $\binom{|N_{p_k}|}{2}-\binom{|N_{p_k}^{\le i}|}{2}$.
\end{proof}

\paragraph{Defining the dynamic program.}
For $j \in \{0,\dots,m\}$, $i \in \{0,\dots,n\}$, and
$q \in \{0,\dots,\binom{n}{2}\}$, let $\mathbf{min}(i,j,q)$ be the minimum cost
of an outcome $P' \subseteq \{p_1,\dots,p_j\}$ such that: (1)
$\pairs(P') \geq q$; (2) the maximum index of any covered voter is exactly $i$
(with $i=0$ for the empty outcome). If no such outcome exists, set
$\mathbf{min}(i,j,q)=\infty$.

We define a DP table $\DP(i,j,q)$ intended to equal $\mathbf{min}(i,j,q)$.
Initialize $\DP(0,0,0)=0$ and $\DP(i,j,q)=\infty$ for all other states.

We fill the table for $j=1,\dots,m$. For each state $(i,j,q)$, we always allow
skipping $p_j$:
\[
\DP(i,j,q) := \min\{\DP(i,j,q),\, \DP(i,j-1,q)\}.
\]
Now consider including $p_j$. Let
$\rho_j := \max\{\ell \in \{1,\dots,n\} : v_\ell \in N_{p_j}\}$ and
$i^+ := \max\{i,\rho_j\}$.  For each $q$ with
$\DP(i,j-1,q) \ne \infty$, define
\[
\Delta_j(i)
:=
\binom{|N_{p_j}|}{2} - \binom{|N_{p_j}^{\le i}|}{2}.
\]
For every $\ell\in\{0,\dots,\Delta_j(i)\}$ with
$q+\ell\le \binom{n}{2}$, we update
\begin{align*}
    \DP(i^+,j,\,q+\ell) :=\;&
    \min\bigl\{\DP(i^+,j,\,q+\ell),\\
    &\DP(i,j-1,q) + \cost(p_j)\bigr\}.
\end{align*}

\paragraph{Correctness.}
We prove by induction on $j$ that after processing projects up to $p_j$, we have
$\DP(i,j,q)=\mathbf{min}(i,j,q)$ for all $i$ and $q$.  The base case is
immediate.

For soundness, every skip transition preserves the represented outcome.  Every
include transition adds $p_j$, and Lemma~\ref{lem:PAIRS0} shows that the true
marginal gain is at least $\Delta_j(i)$; since we update only thresholds
$q+\ell$ with $\ell\le \Delta_j(i)$, the table never claims a threshold that the
corresponding outcome fails to reach.

For completeness, fix a minimum-cost outcome $P'$ witnessing
$\mathbf{min}(i,j,q)$.  If two projects in $P'$ are such that one dominates the
other, deleting the dominated project cannot decrease \pairs{} and strictly
decreases cost.  Thus we may assume that $P'$ contains no dominated project.  If
$p_j\notin P'$, the skip transition and the induction hypothesis apply.  If
$p_j\in P'$, let $P''=P'\setminus\{p_j\}$ and let $i_0$ be the maximum voter
index covered by $P''$.  By Lemma~\ref{lem:PAIRS2}, the marginal gain of adding
$p_j$ to $P''$ is exactly $\Delta_j(i_0)$.  Since
$\pairs(P')\ge q$, we have
$\pairs(P'')\ge \max\{0,q-\Delta_j(i_0)\}$.  The induction hypothesis gives a
predecessor state of cost at most $\cost(P'')$, and the include transition with
the appropriate value of $\ell\le \Delta_j(i_0)$ reaches threshold $q$ with cost
at most $\cost(P')$.

\paragraph{Runtime}
The displayed range of threshold updates can be implemented in linear time
per layer of states. From each predecessor, write only its largest certified
threshold. After processing a project, sweep \(q\) downwards, replacing each
entry by the minimum of itself and the entry at \(q+1\) for the same frontier.
This gives every smaller threshold the same admissible cost. Precompute
\(\Delta_j(i)\) by prefix counts. The table requires \(O(mn^3)\) arithmetic
operations once the representation is available.
\subsection{Proof of Theorem \ref{thm:CONS_poly}}
\begin{proof}
Arrange the projects and voters as in the proof of
Theorem~\ref{thm:PAIRS_poly}.

\paragraph{Calculating the marginal gain in \cons.}
Let $P'$ be an outcome whose maximum project index is $j$, and let $i$ be the maximum index of any voter covered by $P'$. Consider adding a project $p_k$ with $k>j$. Let $N_{p_k}$ denote the set of voters approving $p_k$, and define $N_{p_k}^{\le i} := N_{p_k} \cap \{v_1,\dots,v_i\}$ and $N_{p_k}^{> i} := N_{p_k} \cap \{v_{i+1},\dots,v_n\}$.

Let $N_{P'} := \{v \in N : A_v \cap P' \neq \emptyset\}$ be the set of voters covered by $P'$. Connectivity induces an equivalence relation on $N_{P'}$, yielding a partition into disjoint equivalence classes $(E_1,\dots,E_\gamma)$. Let $E_i$ denote the equivalence class containing voter $v_i$.

When project $p_k$ is added, each voter in $N_{p_k}$ either joins an existing
equivalence class or forms a new one. We first show that if $p_k$ covers both
an already-covered part and a new voter beyond the frontier, then the join is
with the frontier class $E_i$.

\begin{lemma}\label{lem:CONS1}
If $N_{p_k}^{\le i} \neq \emptyset$ and $N_{p_k}^{> i} \neq \emptyset$, then all voters in $N_{p_k}$ belong to the equivalence class $E_i$ after adding $p_k$.
\end{lemma}
\begin{proof}
If $v_i$ approves $p_k$, the claim is immediate.  Suppose that $v_i$ does not
approve $p_k$.  Since $N_{p_k}^{> i}\neq\emptyset$, choose
$u\in N_{p_k}^{> i}$.  The case $R_{p_k}<L_{v_i}$ is impossible: because
$u>i$, we have $L_u\ge L_{v_i}>R_{p_k}$, so $u$ could not approve $p_k$.
Thus we must have $L_{p_k}>R_{v_i}$.

Now choose $v\in N_{p_k}^{\le i}$.  Since $v\le i$, we have
$L_v\le L_{v_i}$; since $v$ approves $p_k$ and $L_{p_k}>R_{v_i}$, we also have
$R_v\ge L_{p_k}>R_{v_i}$.  Therefore the interval of $v$ contains the interval
of $v_i$.  In particular, $v$ approves every project approved by $v_i$, so $v$
already belongs to $E_i$.  Adding $p_k$ connects every voter in $N_{p_k}$ to
$v$, and hence to $E_i$.
\end{proof}

\begin{lemma}\label{lem:CONS_frontier_project}
If $P'$ contains no dominated project, then the component containing the
rightmost project $p_j$ is $E_i$.
\end{lemma}
\begin{proof}
Let $p_t\in P'$ be a project approved by $v_i$.  If $v_i$ approves $p_j$, then
the claim is immediate.  Otherwise, the intervals of $v_i$ and $p_j$ are
disjoint.  The case $R_{p_j}<L_{v_i}$ is impossible by the same argument as in
Lemma~\ref{lem:PAIRS1}: since $t\le j$ and $p_j$ is not dominated by $p_t$, we
would have $R_{p_t}\le R_{p_j}<L_{v_i}$, contradicting that $v_i$ approves
$p_t$.

Hence $L_{p_j}>R_{v_i}$.  Let $v$ be any supporter of $p_j$.  Since $i$ is the
maximum covered voter index, $v\le i$, and therefore $L_v\le L_{v_i}$.  Since
$v$ approves $p_j$, we also have $R_v\ge L_{p_j}>R_{v_i}$.  Thus the interval
of $v$ contains the interval of $v_i$, so $v$ approves $p_t$.  Therefore every
supporter of $p_j$ is connected to $v_i$ through $p_t$, and the component
containing $p_j$ is $E_i$.
\end{proof}

Let $\cons(p_k \mid P') := \cons(P' \cup \{p_k\}) - \cons(P')$ denote the marginal gain.

\begin{lemma}\label{lem:CONS2}
\begin{enumerate}
    \item If $N_{p_k}^{\le i} = \emptyset$, then $\cons(p_k \mid P') = \binom{|N_{p_k}|}{2}$.
    \item If $N_{p_k}^{\le i} \neq \emptyset$ and $N_{p_k}^{> i} \neq \emptyset$, then $\cons(p_k \mid P') \geq \binom{|N_{p_k}^{> i}|}{2} + |N_{p_k}^{> i}| \cdot |E_i|$.
    \item If $P' \cup \{p_k\}$ contains no dominated project and $N_{p_k}^{\le i} \neq \emptyset$, then $\cons(p_k \mid P') = \binom{|N_{p_k}^{> i}|}{2} + |N_{p_k}^{> i}| \cdot |E_i|$.

\end{enumerate}
\end{lemma}

\begin{proof}
In case (1), no voter approving $p_k$ is covered by $P'$, so all voters in $N_{p_k}$ form a new equivalence class, contributing $\binom{|N_{p_k}|}{2}$ new connected pairs.

In case (2), by Lemma~\ref{lem:CONS1}, all voters in $N_{p_k}$ join $E_i$. Since voters in $N_{p_k}^{> i}$ were not previously covered, all pairs between $N_{p_k}^{> i}$ and $E_i$ are newly formed, contributing $|N_{p_k}^{> i}| \cdot |E_i|$ pairs. Additionally, all pairs within $N_{p_k}^{> i}$ are newly connected, contributing $\binom{|N_{p_k}^{> i}|}{2}$ pairs. All pairs already within $E_i$ were previously counted.

In case (3), Lemma~\ref{lem:PAIRS1} implies that all voters in
$N_{p_k}^{\le i}$ approve $p_j$, and Lemma~\ref{lem:CONS_frontier_project}
implies that the component containing $p_j$ is $E_i$. Hence all already-covered
voters approving $p_k$ are already in $E_i$.  The only newly covered voters are
the voters in $N_{p_k}^{> i}$, and they contribute exactly
$\binom{|N_{p_k}^{> i}|}{2}+|N_{p_k}^{> i}|\cdot |E_i|$ new connected pairs.
\end{proof}

\paragraph{Defining the dynamic program.}
Compared to the DP for Theorem~\ref{thm:PAIRS_poly}, we must additionally
track the size of the equivalence class containing the highest-index covered
voter.

For $i \in \{0,\dots,n\}$, $j \in \{0,\dots,m\}$,
$q \in \{0,\dots,\binom{n}{2}\}$, and $w \in \{0,\dots,n\}$, let
$\mathbf{min}(i,j,q,w)$ be the minimum cost of an outcome
$P' \subseteq \{p_1,\dots,p_j\}$ such that: (i) $\cons(P') \ge q$; (ii) the
maximum covered voter index is $i$; and (iii) the equivalence class containing
$v_i$ has size at least $w$ (with $i=w=0$ for the empty outcome). If no such
outcome exists, set $\mathbf{min}(i,j,q,w)=\infty$.

We define a DP table $\DP(i,j,q,w)$ intended to equal $\mathbf{min}(i,j,q,w)$.
Initialize $\DP(0,0,0,0)=0$ and $\DP(i,j,q,w)=\infty$ otherwise.

We fill the table for $j=1,\dots,m$. For each state $(i,j,q,w)$, skipping $p_j$ yields
\[
\DP(i,j,q,w) := \min\{\DP(i,j,q,w),\,\DP(i,j-1,q,w)\}.
\]

Now consider including $p_j$. Let
$\rho_j := \max\{\ell:v_\ell\in N_{p_j}\}$ and
$\ell_j := \min\{\ell:v_\ell\in N_{p_j}\}$.

If $\rho_j\le i$, then the lower-bound contribution of $p_j$ is zero and, because
costs are positive, the transition is never better than skipping $p_j$.  We
therefore only consider states with $\rho_j>i$.
\begin{itemize}
\item If $\ell_j > i$ (a new equivalence class is created), let
\[
\Delta_j(i,w)=\binom{|N_{p_j}|}{2}
\qquad\text{and}\qquad
w' = |N_{p_j}|.
\]
For every $\alpha\in\{0,\dots,\Delta_j(i,w)\}$ with
$q+\alpha\le \binom{n}{2}$, update
\begin{align*}
\DP(\rho_j,j,q+\alpha,w') :=\;&
\min\{\DP(\rho_j,j,q+\alpha,w'),\\
&\DP(i,j-1,q,w)+\cost(p_j)\}.
\end{align*}

\item If $\ell_j \le i$ (the project merges into the frontier class), let
\[
\begin{aligned}
\Delta_j(i,w)&=\binom{|N_{p_j}^{> i}|}{2}+|N_{p_j}^{> i}|\cdot w,\\
w' &= w+|N_{p_j}^{> i}|.
\end{aligned}
\]
For every $\alpha\in\{0,\dots,\Delta_j(i,w)\}$ with
$q+\alpha\le \binom{n}{2}$, update
\begin{align*}
&\DP(\rho_j,j,q+\alpha,w') :=\\\;&
\min\{\DP(\rho_j,j,q+\alpha,w'),\\
&\DP(i,j-1,q,w)+\cost(p_j)\}.
\end{align*}
\end{itemize}
After each layer, propagate each finite cost to all smaller score and
frontier-size thresholds at the same frontier index. It suffices to sweep
each coordinate once in decreasing order.

\paragraph{Correctness.}
The correctness proof follows the same inductive structure as for PAIRS.  For
soundness, Lemma~\ref{lem:CONS2}(1) gives the exact value when $p_j$ starts a
new component, and Lemma~\ref{lem:CONS2}(2) gives a valid lower bound when it
extends the frontier component.  Since the DP updates only objective thresholds
and frontier-class-size thresholds certified by these lower bounds, every table
entry corresponds to an outcome that really attains the stored values.

For completeness, take a minimum-cost witness for
$\mathbf{min}(i,j,q,w)$ and delete any selected project that is dominated by
another selected project; this cannot decrease \cons{} and only lowers the
cost.  Hence a minimum-cost witness can be assumed to contain no dominated
project.  If the witness includes $p_j$, Lemma~\ref{lem:CONS2}(3) gives the
exact marginal contribution and the transition above updates the
frontier-class-size threshold to the exact new value, either $|N_{p_j}|$ for a
new component or $w+|N_{p_j}^{>i}|$ when the frontier component is extended.
The induction hypothesis then supplies the predecessor state, completing the
proof.

\paragraph{Runtime.}
For each predecessor, write the largest certified \(q\) and \(w\), then use
the two decreasing sweeps above. With the supporter prefix counts
precomputed, the table uses \(O(mn^4)\) arithmetic operations. All stored
costs are sums of input costs and have polynomial bit length.

\end{proof}

\subsection{Proof of Theorem \ref{thm:FPT_PAIRS}}
Let \(D\subseteq P\) be the supplied deletion set and \(d=|D|\).
For each of its \(2^d\) subsets \(D'\), select exactly \(D'\) from \(D\).
Discard the branch if \(\cost(D')>b\). On the VCI profile restricted to
\(P\setminus D\), the residual budget is \(b-\cost(D')\). Write
\[
\mathit{cov}_{D'}=\{\{u,v\}\in N^{(2)}:
 A_u\cap A_v\cap D'\ne\emptyset\}.
\]
For \(W\subseteq P\setminus D\), let \(\pairs_{-D'}(W)\) count the unordered
pairs outside \(\mathit{cov}_{D'}\) whose two voters approve a common project
in \(W\). Then
\[
\pairs(D'\cup W)=|\mathit{cov}_{D'}|+\pairs_{-D'}(W).
\]

\paragraph{Calculating the additional PAIRS excluding $\mathit{cov}_{D'}$.}
Let $P'$ be an outcome whose maximum project index is $j$, and let $i$ be the
maximum index of any voter covered by $P'$.
Suppose we add a project $p_k$ with $k > j$.
Let $N_{p_k}^{\le i} := N_{p_k} \cap \{v_1,\dots,v_i\}$ and
$N_{p_k}^{> i} := N_{p_k} \cap \{v_{i+1},\dots,v_n\}$.  Define the set of
candidate pairs newly exposed by $p_k$ as
\[
R_i(p_k)
:=
\{\{u,v\}\subseteq N_{p_k} : \{u,v\}\cap N_{p_k}^{>i}\neq\emptyset\}.
\]
These are precisely the pairs counted by the VCI marginal-gain formula before
accounting for pairs that were already covered by $D'$.

The following lemma characterizes the marginal contribution of $p_k$
under the modified objective.

\begin{lemma}\label{lem:PAIRS3}
If $P' \cup \{p_k\}$ does not contain a dominated project, then
\[
\pairs_{-D'}(p_k \mid P')
=
|R_i(p_k)\setminus \mathit{cov}_{D'}|.
\]
\end{lemma}

\paragraph{Adaptation of the dynamic program.}
Use the initialization and skip transition of
Theorem~\ref{thm:PAIRS_poly}. When taking \(p_j\), use frontier
\(\max\{i,\rho_j\}\) and gain
\(\Delta_j(i)=|R_i(p_j)\setminus\mathit{cov}_{D'}|\).
Each credited pair contains a previously uncovered voter and is outside
\(\mathit{cov}_{D'}\), which proves soundness. On a domination-free witness,
Lemma~\ref{lem:PAIRS2} identifies exactly the newly covered pairs as
\(R_i(p_j)\). Removing the pairs in \(\mathit{cov}_{D'}\) proves
Lemma~\ref{lem:PAIRS3} and gives completeness by the same induction.
Let \(q^*\) be the largest threshold attained within the residual budget.
The branch succeeds exactly when \(q^*+|\mathit{cov}_{D'}|\ge\kappa\).
Constructing the pair sets and filling the table takes polynomial time per
branch, giving total time \(2^d\operatorname{poly}(|I|)\).

\subsection{Proof of Theorem \ref{thm:XP_CONS}}
\begin{proof}
Let $I=(P,A,b,c)$ be an instance and let $D\subseteq P$ be a supplied
deletion set of size $d$ such that the projects in $Q:=P\setminus D$ admit a
VCI representation.  Write $L_z=x_z-r_z$ and $R_z=x_z+r_z$ for every voter
or project in this representation.  Order the projects in $Q$ as
$p_1,\ldots,p_m$ by nondecreasing centre and the voters as
$v_1,\ldots,v_n$ by nondecreasing left endpoint $L_v$.  Projects in $Q$ with
no supporters may be discarded.  For every remaining $p\in Q$, define
\[
    \rho(p):=\max\{h:v_h\in N_p\}.
\]

We enumerate $D'\subseteq D$ and condition on selecting exactly $D'$ from
$D$.  Branches with $\cost(D')>b$ are discarded.  Put
\[
    X:=N_{D'}=\{v\in N:A_v\cap D'\ne\emptyset\},
\]
and let $\mathcal E=\{X_1,\ldots,X_a\}$ be the voter components induced by
$D'$.  These sets, including their members, are computed at the beginning of
the branch.  Notice that $a\le |D'|\le d$.  We call a component containing one
of the sets $X_h$ \emph{anchored}.

We may restrict attention to outcomes in which the selected projects from
$Q$ form an antichain under interval containment.  Indeed, if the interval of
one selected VCI project contains that of another, the former has a superset
of the latter's supporters.  Removing the latter therefore removes no edge
from the voter-connectivity graph and does not increase the cost.  The same
argument lets us retain at most one of two projects having the same interval.
Consequently, in their order by centre, the left and right endpoints of the
selected VCI projects are both strictly increasing.  To see this, for two
intervals with increasing centres, failure of either endpoint to increase
makes one of the intervals contain the other.  It is therefore enough to
remember the last selected VCI project in order to enforce the antichain
condition.

The following frontier observation is the key to the proof.  The argument uses
one global VCI frontier together with the explicitly known exceptional voters
in each anchored component.

\begin{lemma}\label{lem:XP_CONS_frontier}
Let $S\subseteq\{p_1,\ldots,p_{j-1}\}$ be a nonempty antichain, let $p_t$ be
its rightmost project, and let
\[
    i=\max\{h:v_h\in N_S\}.
\]
Let $F$ be the component containing $p_t$ in the outcome $D'\cup S$.  For a
project $p_j$ that can extend the antichain, the following statements hold.
\begin{enumerate}
    \item Every voter in $N_{p_j}\cap\{v_1,\ldots,v_i\}$ belongs to $F$.
    \item Every component other than $F$ meets $N_{p_j}$ only in $X$.
    \item The voters of $N_{p_j}$ that are already covered by $D'\cup S$ are
    exactly
    \[
        N_{p_j}\cap\bigl(X\cup\{v_1,\ldots,v_i\}\bigr).
    \]
\end{enumerate}
\end{lemma}
\begin{proof}
By Lemma~\ref{lem:PAIRS1}, every voter in
$N_{p_j}\cap\{v_1,\ldots,v_i\}$ also approves $p_t$; hence all such voters
belong to $F$.  This proves (1).  If a component other than $F$ met $p_j$ in
a voter $v\notin X$, then $v$ was covered by a project of $S$, so its index is
at most $i$.  Part (1) would put $v$ in $F$, a contradiction.  This proves
(2).

It remains to prove (3).  A supporter of $p_j$ in $X$ is already covered by
$D'$.  A supporter among $v_1,\ldots,v_i$ approves $p_t$ by (1), and hence is
also already covered.  Conversely, a voter outside $X$ that was previously
covered is covered by $S$ and therefore has index at most $i$.  Intersecting
these statements with $N_{p_j}$ gives the claimed equality.
\end{proof}

\paragraph{States.}
After a prefix of $Q$ has been processed, a state consists of the following
data.
\begin{enumerate}
    \item A coarsening $\pi$ of $\mathcal E$.  A block $C\in\pi$ records the
    initial anchored sets that have been merged into one current component.
    Its fixed set of exceptional voters is
    $X_C:=\bigcup_{X_h\in C}X_h$.
    \item For each $C\in\pi$, the exact current component size $s_C$.
    \item The current VCI-frontier label $\phi$.  It is either a block of
    $\pi$, the symbol $\mathsf U$ for an unanchored component, or
    $\bot$ if no VCI project has been selected.  If
    $\phi=\mathsf U$, the state also stores its size $s_{\mathsf U}$.
    \item The largest voter index $i$ covered by a selected VCI project, the
    last selected VCI project $p_t$, and the exact current value $q$ of
    \cons{}.
\end{enumerate}
For every state the table stores the minimum cost of the selected projects
from $Q$ that realize it.  Components that are neither anchored nor the
current VCI frontier need not be stored.  Indeed, by
Lemma~\ref{lem:XP_CONS_frontier}(2), such a component cannot meet a later
project and is therefore permanently closed.

The initial state has partition $\mathcal E$, sizes $s_{X_h}=|X_h|$,
$\phi=\bot$, $i=t=0$, and
\[
    q_0=\sum_{h=1}^a\binom{|X_h|}{2}.
\]
Its table value is zero; the fixed cost $\cost(D')$ is accounted for when the
branch is tested against the budget.

\paragraph{Transitions.}
Skipping the next project leaves the state unchanged.  Suppose that we include
$p_j$.  If $t>0$, we first check explicitly that
$L_{p_t}<L_{p_j}$ and $R_{p_t}<R_{p_j}$, so that adding $p_j$ preserves the
antichain.  This comparison with the last selected project suffices because
the endpoints of all previously selected projects are increasing.  Let
\[
    Y_i:=X\cup\{v_1,\ldots,v_i\}
    \qquad\text{and}\qquad
    z:=|N_{p_j}\setminus Y_i|.
\]
By Lemma~\ref{lem:XP_CONS_frontier}(3), $z$ is exactly the number of newly
covered voters.

We can also identify exactly the old components joined by $p_j$.  Include an
anchored block $C$ in $\mathcal J$ precisely when
$N_{p_j}\cap X_C\ne\emptyset$.  In addition, include the frontier component
when $\phi\ne\bot$ and
\[
    N_{p_j}\cap\{v_1,\ldots,v_i\}\ne\emptyset.
\]
If the frontier is anchored, duplicates are removed from $\mathcal J$.
Lemma~\ref{lem:XP_CONS_frontier}(1)--(2) shows that this is exactly the set of
old components met by $p_j$: a non-frontier anchored component can be met only
through its explicitly known set $X_C$, while all nonexceptional old
supporters lie in the frontier component.

Let
\[
    s=z+\sum_{H\in\mathcal J}s_H,
\]
where $s_H$ means $s_C$ for an anchored component and $s_{\mathsf U}$ for an
unanchored frontier.  (If $\mathcal J=\emptyset$, this simply gives
$s=|N_{p_j}|$.)  The exact increase in the objective is
\[
    \binom{s}{2}-\sum_{H\in\mathcal J}\binom{s_H}{2}.
\]
All anchored blocks in $\mathcal J$ are replaced by their union and assigned
size $s$.  If $\mathcal J$ contains an anchored block, this union is the new
frontier label; otherwise the new label is $\mathsf U$ with size $s$.
Anchored blocks outside $\mathcal J$ and their sizes are unchanged.  A former
unanchored frontier outside $\mathcal J$ is closed and forgotten.  Finally,
we set $i:=\max\{i,\rho(p_j)\}$, $t:=j$, add the displayed increment to $q$, and
add $c(p_j)$ to the table value.

All quantities used by this transition are determined by the state.  In
particular, the overlap with previously covered voters is
\[
    N_{p_j}\cap N_{D'\cup S}
    =N_{p_j}\cap Y_i.
\]

\paragraph{Correctness.}
We induct on the processed prefix.  The initial state is exactly the component
structure and objective value induced by $D'$.  For the inductive step, a
partial outcome either skips or includes the next project.  In the latter
case, Lemma~\ref{lem:XP_CONS_frontier} identifies both the already-covered
supporters and every old component joined by the project.  Thus the transition
computes the exact new component size and the exact change in \cons{}.  It
also preserves all anchored components and the unique component that can be
met later through a nonexceptional voter.  Conversely, each transition is
realized by adding its project to an outcome realizing the predecessor state.
The table therefore stores the minimum cost for every realizable state.
Moreover, the transition rule depends only on the stored data: the fixed sets
$X_C$ determine all non-frontier intersections, while $i$ determines all
intersections through nonexceptional voters.  Hence retaining only the
minimum-cost realization of a state cannot discard a better continuation.

Every outcome considered by the DP is an antichain.  Conversely, as observed
above, some minimum-cost outcome attaining any feasible \cons{} threshold has
this form.  Hence the original branch is a yes-instance precisely when a final
state with $q\ge\kappa$ has table value at most $b-\cost(D')$.

\paragraph{Runtime.}
For a fixed $D'$, the initial partition has at most $d$ blocks.  It has at most
$B_d=2^{O(d\log d)}$ coarsenings, and for each coarsening there are at most
$n^d$ choices for the component sizes.  The frontier label has at most $d+2$
choices, and $i$, $t$, and $q$ contribute only a polynomial factor.  Thus the
number of states is
\[
    2^{O(d\log d)}n^{d+O(1)}m^{O(1)}.
\]
Each transition is polynomial-time computable using the precomputed voter sets
$X_C$ and $N_{p_j}$.  Multiplying by the $2^d$ choices of $D'$ gives running
time $n^{f(d)}m^{O(1)}$ for a computable function $f$.  This is XP with
respect to $\delta_{\mathrm{VCI}}(I)$.
\end{proof}

\section{Omitted Proofs for Bicriteria Complexity}\label{app:bicriteria-proofs}

\subsection{Proof of Theorem~\ref{thm:theta_2^P}}

\paragraph{Membership.}
For \(f\in\{\pairs,\cons,\av,\cc\}\), an NP oracle can decide whether
some feasible outcome has \(f(W)\ge t\). The outcome is a certificate and
its cost and score are computable in polynomial time. All four scores are
polynomially bounded integers. Binary search therefore finds both optima
with \(O(\log |I|)\) NP queries. One final query asks whether any feasible
outcome simultaneously meets the two resulting score thresholds.
Thus, \OurProblem{} belongs to $\Theta_2^P$.

\paragraph{Hardness.}
We first introduce a promise version of comparing maximum-coverage values.

\begin{tcolorbox}[title=Promise Compare Set Cover (\textnormal{\textsc{Comp-SC}})]
\textbf{Input:}
Two Set Cover instances
\[
I_1=(U_1,\mathcal{F}_1,k_1)
\quad\text{and}\quad
I_2=(U_2,\mathcal{F}_2,k_2),
\]
where $\mathcal{F}_i\subseteq 2^{U_i}$ and $k_i\in\mathbb{N}$.
As usual, we assume that $\bigcup_{S\in\mathcal{F}_i}S=U_i$.

\medskip
\textbf{Promise:}
For each $i\in\{1,2\}$, no subfamily of fewer than $k_i$ sets covers $U_i$.

\tcblower
\textbf{Question:}
Is
\[
\max_{\substack{\mathcal{S}\subseteq\mathcal{F}_1\\|\mathcal{S}|=k_1}}
\left|\bigcup_{S\in\mathcal{S}}S\right|
\geq
\max_{\substack{\mathcal{S}\subseteq\mathcal{F}_2\\|\mathcal{S}|=k_2}}
\left|\bigcup_{S\in\mathcal{S}}S\right|?
\]
\end{tcolorbox}

\begin{lemma}\label{lem:comp-sc-hard}
\textnormal{\textsc{Comp-SC}} is $\Theta_2^P$-hard.
\end{lemma}
\begin{proof}
We reduce from \textsc{Compare-MaxSAT}. For a CNF formula $\varphi$, let
$g(\varphi)$ be the maximum number of simultaneously satisfiable clauses.
We may assume that the two input formulas
$\varphi_i=(X_i,\mathcal{C}_i)$ have equally many variables and clauses.
Indeed, unused variables equalize the numbers of variables. To equalize the
numbers of clauses, suppose, without loss of generality, that
$|\mathcal{C}_1|<|\mathcal{C}_2|$, and let
$q=|\mathcal{C}_2|-|\mathcal{C}_1|$. Introduce fresh variables
$z_1,\ldots,z_q$ in both formulas. For every $j\in[q]$, add the two clauses
$(z_j)$ and $(\neg z_j)$ to $\varphi_1$, and add the clause $(z_j)$ to
$\varphi_2$. The resulting formulas have the same numbers of variables and
clauses, and both optimum values increase by exactly $q$. The comparison is
therefore preserved.

Let $M=|\mathcal{C}_1|+1=|\mathcal{C}_2|+1$. For each $i\in\{1,2\}$, construct
a Set Cover instance with
\[
U_i=\mathcal{C}_i\cup
\bigcup_{x\in X_i}\{(x,j)\mid j\in[M]\}
\quad\text{and}\quad
k_i=|X_i|.
\]
For every $x\in X_i$, introduce the two sets
\begin{align*}
T_x&=\{(x,j)\mid j\in[M]\}
 \cup\{C\in\mathcal{C}_i\mid x\in C\},\\
F_x&=\{(x,j)\mid j\in[M]\}
 \cup\{C\in\mathcal{C}_i\mid \neg x\in C\},
\end{align*}
and let $\mathcal{F}_i=\{T_x,F_x\mid x\in X_i\}$.

Any subfamily of fewer than $k_i$ sets misses the entire $M$-element block
associated with some variable, so the promise is satisfied. We next claim
that an optimal subfamily of size $k_i$ contains exactly one of $T_x,F_x$ for
each $x\in X_i$. Otherwise, it omits both sets for some variable $x$ and
contains both sets for another variable $y$. Removing either $T_y$ or $F_y$
loses at most $|\mathcal{C}_i|=M-1$ clause elements, whereas adding either
$T_x$ or $F_x$ gains the $M$ previously uncovered elements associated with
$x$. This strictly improves the solution.

Thus, every optimal subfamily corresponds to a truth assignment, and
\[
\max_{\substack{\mathcal{S}\subseteq\mathcal{F}_i\\|\mathcal{S}|=k_i}}
\left|\bigcup_{S\in\mathcal{S}}S\right|
=M|X_i|+g(\varphi_i).
\]
Because the two formulas have equally many variables, comparing these two
coverage values is equivalent to comparing $g(\varphi_1)$ and
$g(\varphi_2)$.
\end{proof}

For a Set Cover instance $I=(U,\mathcal{F},k)$, write
\[
f(I)=
\max_{\substack{\mathcal{S}\subseteq\mathcal{F}\\|\mathcal{S}|=k}}
\left|\bigcup_{S\in\mathcal{S}}S\right|.
\]

\begin{lemma}\label{lem:theta_1}
\textnormal{\textsc{Comp-SC}} remains $\Theta_2^P$-hard under the additional promise that
$f(I_1)\neq f(I_2)$.
\end{lemma}
\begin{proof}
Replace every element of each universe by two copies. For the first instance,
also add a new element $u^\star$ to every set. Formally,
\[
\begin{aligned}
U_1'&=(\{1,2\}\times U_1)\cup\{u^\star\},\\
\mathcal{F}_1'&=
\bigl\{\{u^\star\}\cup(\{1,2\}\times S)\mid S\in\mathcal{F}_1\bigr\},
\end{aligned}
\]
and
\[
U_2'=\{1,2\}\times U_2,
\qquad
\mathcal{F}_2'=
\bigl\{\{1,2\}\times S\mid S\in\mathcal{F}_2\bigr\}.
\]
Keeping $k_1$ and $k_2$ unchanged gives
\[
f(I_1')=2f(I_1)+1
\qquad\text{and}\qquad
f(I_2')=2f(I_2).
\]
Hence $f(I_1)\geq f(I_2)$ if and only if $f(I_1')\geq f(I_2')$, and the two
new optimum values are unequal. Any cover of a copied universe projects to a
cover of the original universe, so the original promise is preserved.
\end{proof}

Fix an instance $(I_1,I_2)$ satisfying Lemma~\ref{lem:theta_1}, where
$I_i=(U_i,\mathcal{F}_i,k_i)$. Define
\[
\begin{aligned}
X&=2\left(1+\binom{\max\{|U_1|,|U_2|\}}{2}\right),\\
Y&=\binom{X+|U_1|}{2}+1,\\
Z&=2|\mathcal{F}_1|(|U_1|+X)+2|\mathcal{F}_2|(|U_2|+Y),\\
a&=(k_2-1)(k_1+1)+1.
\end{aligned}
\]
The factor two in the definition of $X$ ensures that $XY$ is even. We
construct a PB instance with budget
\[
b=(k_1+1)k_2+\frac12.
\]

The voters are partitioned into six sets:
\[
\begin{aligned}
 A&=\{A_u\mid u\in U_1\}, & |C|&=X,\\
 B&=\{B_u\mid u\in U_2\}, & |D|&=Y.
\end{aligned}
\]
The set $E$ contains $XY$ voters partitioned into $XY/2$ pairs
$E_1,\ldots,E_{XY/2}$, and
\[
F=\{F_i\mid i\in[2aZ]\}.
\]
The projects are partitioned as follows:
\[
P_1=\{p_S^1\mid S\in\mathcal{F}_1\},\qquad
P_2=\{p_S^2\mid S\in\mathcal{F}_2\},
\]
\[
P_3=\{r_i\mid i\in[XY/2]\},\qquad
P_4=\{d_i\mid i\in[2aZ]\}.
\]
Their costs and supporter sets are
\begin{align*}
\cost(p_S^1)&=1,
&N_{p_S^1}&=C\cup\{A_u\mid u\in S\},\\
\cost(p_S^2)&=k_1+1,
&N_{p_S^2}&=D\cup\{B_u\mid u\in S\},\\
\cost(r_i)&=\frac1{XY},
&N_{r_i}&=E_i,\\
\cost(d_i)&=\frac1{2Z},
&N_{d_i}&=\{F_i\}.
\end{align*}
In particular, $\cost(P_3)=1/2$ and $\cost(P_4)=a$.

\begin{lemma}\label{lem:theta_2}
For each $BR\in\{\pairs,\cons\}$, every $BR$-maximizing outcome contains
exactly $k_2$ projects from $P_2$ and no project from $P_1$. Projects from
$P_4$ make no contribution to either bridging objective and may be removed
without changing the objective value.
\end{lemma}
\begin{proof}
Budget feasibility immediately gives $|P'\cap P_2|\leq k_2$. Suppose that a
$BR$-maximizing outcome $P'$ contains at most $k_2-1$ projects from $P_2$.
Let
\[
Q=P'\setminus(P_1\cup P_4).
\]
By the promise on $I_2$, the sets corresponding to $P'\cap P_2$ do not cover
$U_2$. Choose an uncovered element $u$ and a set $S\in\mathcal{F}_2$ with
$u\in S$. Then
\[
\cost(Q)\leq(k_1+1)(k_2-1)+\frac12,
\]
so $Q\cup\{p_S^2\}$ is feasible. Adding $p_S^2$ connects $B_u$ to every
voter in $D$, and therefore increases either bridging objective by at least
$Y$.

Projects in $P_4$ have only one supporter each and contribute nothing to
either objective. Moreover, all connections contributed by $P_1$ lie within
$A\cup C$. Consequently,
\[
BR(P')\leq BR(Q)+\binom{|U_1|+X}{2}.
\]
Since $Y=\binom{|U_1|+X}{2}+1$, the outcome $Q\cup\{p_S^2\}$ has a strictly
larger $BR$-value, a contradiction. Hence $|P'\cap P_2|=k_2$. These projects
cost $(k_1+1)k_2$, leaving only $1/2$ of the budget, whereas every project in
$P_1$ costs one. Thus $P'\cap P_1=\emptyset$. The final assertion follows
from the singleton supporter sets of projects in $P_4$.
\end{proof}

Let $\OPT_{BR}$ be the maximum attainable $BR$-value, and set
\[
L_2=Y\left(f(I_2)+\frac{X+Y-1}{2}\right).
\]

\begin{lemma}\label{lem:theta_3}
For every $BR\in\{\pairs,\cons\}$,
\[
L_2\leq\OPT_{BR}\leq L_2+\binom{|U_2|}{2}.
\]
\end{lemma}
\begin{proof}
Choose $k_2$ projects from $P_2$ whose corresponding sets cover exactly
$f(I_2)$ elements, and select all projects in $P_3$. This outcome is feasible.
The projects in $P_3$ connect $XY/2$ disjoint voter pairs. The selected
$P_2$-projects connect all pairs within $D$ and connect each of the
$f(I_2)$ represented voters in $B$ to every voter in $D$. Thus, for either
bridging objective, the outcome has value at least
\[
\frac{XY}{2}+\binom{Y}{2}+Yf(I_2)=L_2.
\]

For the upper bound, let $P'$ maximize $BR$. By Lemma~\ref{lem:theta_2}, we
may disregard $P_4$, and $P'$ consists of $k_2$ projects from $P_2$ and some
projects from $P_3$. At most $f(I_2)$ voters in $B$ approve a selected
$P_2$-project. Hence
\begin{align*}
BR(P')
&\leq \frac{XY}{2}+\binom{Y+f(I_2)}{2}\\
&=\frac{XY}{2}+\binom{Y}{2}+Yf(I_2)+\binom{f(I_2)}{2}\\
&\leq L_2+\binom{|U_2|}{2}.
\end{align*}
This bound applies to \pairs{} directly and to \cons{} because all represented
voters in $B$ and all voters in $D$ lie in one connected component.
\end{proof}

Let $\OPT_{REP}$ be the maximum attainable $REP$-value for
$REP\in\{\av,\cc\}$, and define
\[
W=XY+2aZ.
\]

\begin{lemma}\label{lem:theta_4}
For every $REP\in\{\av,\cc\}$,
\[
W\leq\OPT_{REP}\leq W+\frac Z2.
\]
\end{lemma}
\begin{proof}
The outcome $P_3\cup P_4$ has cost $a+1/2\leq b$ and represents every voter
in $E\cup F$. Its \av{} and \cc{} values are both $W$, proving the lower
bound.

For every outcome, $\cc(P')\leq\av(P')\leq\av(P)$. The total approval score
of all projects in $P_1\cup P_2$ is at most
\[
|\mathcal{F}_1|(|U_1|+X)+|\mathcal{F}_2|(|U_2|+Y)=\frac Z2.
\]
Furthermore, $\av(P_3)+\av(P_4)=W$. Thus $\OPT_{REP}\leq W+Z/2$.
\end{proof}

Set $\beta=W/(W+Z/2)$.

\begin{lemma}\label{lem:theta_5}
For every $REP\in\{\av,\cc\}$:
\begin{enumerate}
    \item every feasible outcome containing $P_3\cup P_4$ achieves
    $\beta$-$REP$; and
    \item no feasible outcome $P'$ with $|P'\cap P_1|>k_1$ or
    $P'\cap P_2\neq\emptyset$ achieves $\beta$-$REP$.
\end{enumerate}
\end{lemma}
\begin{proof}
If $P_3\cup P_4\subseteq P'$, then $REP(P')\geq W$. By
Lemma~\ref{lem:theta_4}, this implies
$REP(P')/\OPT_{REP}\geq W/(W+Z/2)=\beta$.

For the second statement, either condition implies
\[
\cost(P'\cap(P_1\cup P_2))\geq k_1+1.
\]
The remaining budget is at most
\[
b-(k_1+1)=(k_2-1)(k_1+1)+\frac12=a-\frac12.
\]
Since $\cost(P_4)=a$, projects of total cost at least $1/2$ must be omitted
from $P_4$. Each such project has cost $1/(2Z)$ and is the only project
approved by its corresponding voter in $F$, so at least $Z$ voters in $F$
receive no representation. It follows that
\[
REP(P')\leq \av(P)-Z\leq W-\frac Z2.
\]
Using $\OPT_{REP}\geq W$, we obtain
\[
\frac{REP(P')}{\OPT_{REP}}
\leq\frac{W-Z/2}{W}
<\frac{W}{W+Z/2}=\beta,
\]
where the strict inequality follows by cross multiplication.
\end{proof}

For a fixed $REP\in\{\av,\cc\}$, let $\mathrm{FEAS}_{BR}$ be the maximum
$BR$-value among feasible outcomes that achieve $\beta$-$REP$, and set
\[
L_1=X\left(f(I_1)+\frac{X+Y-1}{2}\right).
\]

\begin{lemma}\label{lem:theta_6}
For every $BR\in\{\pairs,\cons\}$,
\[
L_1\leq\mathrm{FEAS}_{BR}\leq L_1+\binom{|U_1|}{2}.
\]
\end{lemma}
\begin{proof}
Choose $k_1$ projects from $P_1$ whose corresponding sets cover exactly
$f(I_1)$ elements, and add all projects in $P_3\cup P_4$. Its cost is
\[
k_1+\frac12+a=(k_1+1)k_2+\frac12=b,
\]
and it achieves $\beta$-$REP$ by Lemma~\ref{lem:theta_5}. Its bridging value
is at least
\[
\frac{XY}{2}+\binom{X}{2}+Xf(I_1)=L_1.
\]

Conversely, let $P'$ be a $\beta$-$REP$ outcome maximizing $BR$.
Lemma~\ref{lem:theta_5} gives $P'\cap P_2=\emptyset$ and
$|P'\cap P_1|\leq k_1$. Thus, at most $f(I_1)$ voters in $A$ approve a
selected project from $P_1$. Projects in $P_4$ contribute no bridging value,
and projects in $P_3$ contribute at most $XY/2$. Therefore,
\begin{align*}
BR(P')
&\leq\frac{XY}{2}+\binom{X+f(I_1)}{2}\\
&=\frac{XY}{2}+\binom{X}{2}+Xf(I_1)+\binom{f(I_1)}{2}\\
&\leq L_1+\binom{|U_1|}{2}.
\end{align*}
\end{proof}

We complete the reduction by setting $\alpha=X/Y$. The constructed instance
is a yes-instance of \OurProblem{} if and only if
\[
\frac{\mathrm{FEAS}_{BR}}{\OPT_{BR}}\geq\frac XY.
\]
If $f(I_1)\geq f(I_2)$, then the unequal-optima promise gives
$f(I_1)\geq f(I_2)+1$. Lemmas~\ref{lem:theta_3} and~\ref{lem:theta_6}, and
$Y>\binom{|U_2|}{2}$, imply
\[
\frac{\mathrm{FEAS}_{BR}}{\OPT_{BR}}
\geq
\frac{X\left(f(I_1)+\frac{X+Y-1}{2}\right)}
{Y\left(f(I_2)+\frac{X+Y-1}{2}\right)+\binom{|U_2|}{2}}
\geq\frac XY.
\]
If $f(I_1)<f(I_2)$, then $f(I_2)\geq f(I_1)+1$. Since
$X>\binom{|U_1|}{2}$, the same lemmas give
\[
\frac{\mathrm{FEAS}_{BR}}{\OPT_{BR}}
\leq
\frac{X\left(f(I_1)+\frac{X+Y-1}{2}\right)+\binom{|U_1|}{2}}
{Y\left(f(I_2)+\frac{X+Y-1}{2}\right)}
<\frac XY.
\]
The construction works for every pair
$(BR,REP)\in\{\pairs,\cons\}\times\{\av,\cc\}$, proving
$\Theta_2^P$-hardness.

\subsection{Proof of Theorem~\ref{thm:sigma_2^P}}

\paragraph{Membership.}
Guess a feasible outcome $P'$. Whether $P'$ fails to achieve
$\alpha$-$BR$ has an NP certificate: a feasible outcome whose $BR$-value is
larger than $BR(P')/\alpha$ (with the case $\alpha=0$ being immediate).
Likewise, failure of $\beta$-EJR has an NP
certificate consisting of a voter coalition and a commonly approved project
set witnessing an EJR violation. An NP machine with an NP oracle can therefore
verify both required properties, so \OurProblem{} belongs to $\Sigma_2^P$.

\paragraph{Hardness.}
We reduce from the $\Sigma_2^P$-complete problem \clique{}
\citep{rutenburg1994propositional,grune2025completeness}.

\begin{tcolorbox}[title=\textsc{Min--Max Clique Deletion} (\clique)]
\textbf{Input:}
A graph $G=(V,E)$ and integers $k,\ell\in\mathbb{N}_{>0}$.

\tcblower
\textbf{Question:}
Is there a set $V'\subseteq V$ with $|V'|=\ell$ such that $G-V'$ contains
no clique of size $k$?
\end{tcolorbox}

Let $(G,k,\ell)$ be an instance of \clique{}, and let
\[
d=\max_{v\in V}\deg(v),
\qquad
q=|V|-k-1.
\]
By adding isolated vertices, we may assume that $|V|\geq4$, $d\geq1$,
$|E|<|V|-k$, and $|V|-k-\binom{k}{2}>0$. These assumptions do not change
the existence of a deletion set of size $\ell$; trivial exceptional instances
can be decided directly.

Construct the following five types of voters:
\begin{align*}
A&=\{A_e\mid e\in E\},&
B&=\{B_i\mid v_i\in V\},\\
C&=\bigcup_{v_i\in V}C_i,
&|C_i|&=d-\deg(v_i),\\
|D|&=|V|-k-\binom{k}{2},&
F&=F_1\cup F_2\cup F_3,
\end{align*}
where the sets $F_1,F_2,F_3$ are disjoint and
$|F_j|=|V|^3$ for each $j\in[3]$. Let $N=A\cup B\cup C\cup D\cup F$ and
$n=|N|$. Finally, define
\[
\epsilon=\frac{\ell q}{2n-3q}.
\]
The denominator is positive, and the padding ensures that $\epsilon<1/3$.
The budget is $b=\ell+3\epsilon$.

The fixed class is \(P_1=\{p_1,p_2,p_3\}\). Each vertex \(v_i\) indexes
one project \(d_i\) in \(P_2\) and one project \(r_i\) in \(P_3\).
For each $v_i\in V$, let
\[
H_i=\{A_e\in A\mid e\text{ is incident to }v_i\}.
\]
The costs and supporter sets are
\begin{align*}
\cost(p_1)&=\epsilon,
&N_{p_1}&=A\cup C\cup D\cup F_1,\\
\cost(p_2)&=\epsilon,
&N_{p_2}&=A\cup F_2,\\
\cost(p_3)&=\epsilon,
&N_{p_3}&=D\cup F_3,\\
\cost(d_i)&=1,
&N_{d_i}&=H_i\cup C_i\cup\{B_i\},\\
\cost(r_i)&=\frac{\epsilon}{q},
&N_{r_i}&=D\cup(A\setminus H_i).
\end{align*}

\begin{lemma}\label{lem:sigma_2^P}
For either $BR\in\{\pairs,\cons\}$, an outcome $P'$ maximizes $BR$ if and
only if
\[
P_1\subseteq P',\qquad P'\cap P_3=\emptyset,
\qquad\text{and}\qquad |P'\cap P_2|=\ell.
\]
In particular, all such outcomes have the same \pairs{} value and the same
\cons{} value.
\end{lemma}
\begin{proof}
Let $N_0=A\cup B\cup C\cup D$. Since $|C|\leq |V|^2$ and
$|A|+|B|+|D|<3|V|$, we have $|N_0|<|V|^3/2$.

We first show that every maximizing outcome contains $P_1$. Fix
$p_j\in P_1$. Projects in $P_2\cup P_3$ have no supporters in $F$, so their
marginal \pairs{} contribution relative to $P_1\setminus\{p_j\}$ is at most
$\binom{|N_0|}{2}$. In contrast, adding $p_j$ connects every pair among the
$|V|^3$ voters in $F_j$. Hence
\[
\begin{aligned}
\pairs(P\setminus\{p_j\})
&\leq \pairs(P_1\setminus\{p_j\})+\binom{|N_0|}{2}\\
&<\pairs(P_1).
\end{aligned}
\]
For \cons{}, an outcome omitting $p_j$ leaves all voters in $F_j$
unrepresented, and thus
\[
\begin{aligned}
\cons(P\setminus\{p_j\})
&\leq\binom{2|V|^3+|N_0|}{2}\\
&<\binom{3|V|^3}{2}\leq\cons(P_1).
\end{aligned}
\]
Therefore, omitting any project in $P_1$ precludes optimality.

Next observe that $N_{r_i}\subseteq N_{p_1}$ for every $r_i\in P_3$.
Consequently, once $p_1$ is selected, projects in $P_3$ add no \pairs{} or
\cons{} connections. If a feasible outcome containing $P_1$ also contains a
project from $P_3$, then it contains at most $\ell-1$ unit-cost projects from
$P_2$. Remove all projects in $P_3$ and add an omitted project $d_i$.
The resulting outcome is feasible. Since $B_i$ approves no other project and
$|H_i|+|C_i|=d\geq1$, adding $d_i$ strictly increases both bridging
objectives. Thus, no maximizing outcome contains a project from $P_3$.

It remains to determine the choice from $P_2$. After selecting $P_1$, the
remaining budget is exactly $\ell$, so at most $\ell$ projects from $P_2$ can
be selected. Every additional $d_i$ strictly increases both objectives, and
therefore exactly $\ell$ are selected.

Finally, let $S\subseteq P_2$. Because $H_i\cup C_i\subseteq N_{p_1}$ and
$B_i$ approves only $d_i$, the marginal \pairs{} contribution of every
$d_i$ relative to $P_1\cup(S\setminus\{d_i\})$ is
$|H_i|+|C_i|=d$. Hence
\[
\pairs(P_1\cup S)=\pairs(P_1)+d|S|.
\]
The projects in $P_1$ place all voters in $A\cup C\cup D\cup F$ in one
\cons{} component: $p_1$ overlaps with $p_2$ on the nonempty set $A$ and
with $p_3$ on the nonempty set $D$. Each selected $d_i$ then adds precisely
$B_i$ to that component. Therefore,
\[
\cons(P_1\cup S)
=\binom{|A|+|C|+|D|+|F|+|S|}{2}.
\]
Both values depend only on $|S|$, which proves the claim.
\end{proof}

For an outcome \(P'\), an \emph{EJR violation} consists of a nonempty
coalition \(S\) and a bundle \(T\) approved by all its members, with
\(\cost(T)\le |S|b/n\) and \(\cost(P'\cap A_i)<\cost(T)\) for every
\(i\in S\).

\begin{lemma}\label{lem:ejr-violation}
Let $S\subseteq N$ and $T\subseteq\bigcap_{i\in S}A_i$.
\begin{enumerate}
    \item If $(S,T)$ is an EJR violation and $|S|\leq q$, then
    $\cost(P'\cap A_i)<2\epsilon$ for every $i\in S$.
    \item If $(S,T)$ is an EJR violation and $|S|\leq2|V|$, then
    $\cost(P'\cap A_i)<1$ for every $i\in S$.
    \item If $(S,T)$ is an EJR violation and $\cost(T)\geq1$, then
    $|S|>2|V|$.
    \item If $(S,T)$ is an EJR violation and
    $\cost(P'\cap A_i)=2\epsilon$ for every $i\in S$, then
    $\cost(T)>2\epsilon$.
    \item If $|S|\geq q+1$, $\cost(T)=2\epsilon+\epsilon/q$, and
    $\cost(P'\cap A_i)=2\epsilon$ for every $i\in S$, then $(S,T)$ is an
    EJR violation.
\end{enumerate}
\end{lemma}
\begin{proof}
By the definition of $\epsilon$, the proportional budget share of one voter
is
\[
\frac{b}{n}
=\frac{\ell+3\epsilon}{n}
=\frac{2\epsilon}{q}.
\]
Statement 1 follows because a coalition of at most $q$ voters can afford
projects of total cost at most $2\epsilon$.

For a coalition of at most $2|V|$ voters, the proportional share is at most
\[
\frac{4|V|\epsilon}{q}
=\frac{4|V|\ell}{2n-3q}
<1,
\]
where we use $\ell\leq|V|$ and $n>3|V|^3$. This proves Statement 2, and
Statement 3 is its contrapositive. Statement 4 follows immediately from the
strict inequality in the definition of an EJR violation.

Finally, a coalition of at least $q+1$ voters has proportional share at
least
\[
(q+1)\frac{2\epsilon}{q}
=2\epsilon+\frac{2\epsilon}{q}
>2\epsilon+\frac{\epsilon}{q}.
\]
Together with common approval and
$\cost(P'\cap A_i)=2\epsilon<\cost(T)$ for every $i\in S$, this proves
Statement 5.
\end{proof}

We now prove that $(G,k,\ell)$ is a yes-instance of \clique{} if and only if
the constructed PB instance admits an outcome that simultaneously maximizes
$BR\in\{\pairs,\cons\}$ and satisfies EJR.

\paragraph{Forward direction.}
Suppose that some $V'\subseteq V$ with $|V'|=\ell$ intersects every
$k$-clique. Consider
\[
P'=P_1\cup\{d_i\in P_2\mid v_i\in V'\}.
\]
This outcome has cost $\ell+3\epsilon=b$ and maximizes both bridging
objectives by Lemma~\ref{lem:sigma_2^P}. We show that it satisfies EJR.

Suppose, toward a contradiction, that $(S,T)$ is an EJR violation. First,
$T\cap P_2=\emptyset$. Indeed, if $d_i\in T$, then $\cost(T)\geq1$, so
Lemma~\ref{lem:ejr-violation}(3) gives $|S|>2|V|$. On the other hand, common
approval implies $S\subseteq N_{d_i}$, while
$|N_{d_i}|=d+1\leq|V|$, a contradiction.

Next, $S$ contains no voter from $B\cup C\cup F$. Every such voter approves
only projects in $P_1\cup P_2$. Since $T\cap P_2=\emptyset$, including such a
voter in $S$ would imply $T\subseteq P_1\subseteq P'$, and hence
$\cost(P'\cap A_i)\geq\cost(T)$ for that voter, contradicting the definition
of a violation. Therefore, $S\subseteq A\cup D$.

Every voter in $A\cup D$ receives projects of cost at least $2\epsilon$ from
$P_1$. Lemma~\ref{lem:ejr-violation}(1) therefore implies
$|S|\geq q+1=|V|-k$. Since $|A|=|E|<|V|-k$ and
$|D|=|V|-k-\binom{k}{2}<|V|-k$, the coalition $S$ intersects both $A$ and
$D$. Voters in $D$ disapprove $p_2$, and voters in $A$ disapprove $p_3$.
Consequently,
\[
T\subseteq\{p_1\}\cup P_3.
\]

Let $A'=S\cap A$. Then $|A'|\geq\binom{k}{2}$. If an edge $e$ represented
by a voter in $A'$ were incident to a vertex in $V'$, that voter would
receive a selected unit-cost project $d_i$. Since
$|S|\leq|A|+|D|<2|V|$, this would contradict
Lemma~\ref{lem:ejr-violation}(2). Thus, all edges represented in $A'$ belong
to $G-V'$.

Let $V_{A'}$ be the set of endpoints of the edges represented in $A'$.
A simple graph on fewer than $k$ vertices has fewer than $\binom{k}{2}$
edges, so $|V_{A'}|\geq k$. Equality would imply that these $k$ vertices form
a clique, which is impossible in $G-V'$. Hence $|V_{A'}|\geq k+1$.
For every $v_i\in V_{A'}$, some voter in $A'$ corresponds to an edge
incident to $v_i$ and therefore disapproves $r_i$. Common approval gives
$r_i\notin T$. It follows that $T$ contains at most
$|V|-(k+1)=q$ projects from $P_3$, and hence
\[
\cost(T)\leq\epsilon+q\frac{\epsilon}{q}=2\epsilon.
\]
All voters in $S$ receive exactly $2\epsilon$ from $P'$: voters in $D$
receive $p_1,p_3$, while voters in $A'$ receive $p_1,p_2$ and no selected
incident $d_i$. This contradicts Lemma~\ref{lem:ejr-violation}(4). Therefore,
$P'$ satisfies EJR.

\paragraph{Reverse direction.}
Suppose that an outcome $P'$ maximizes $BR$ and satisfies EJR.
Lemma~\ref{lem:sigma_2^P} gives
\[
P'=P_1\cup\{d_i\mid v_i\in V'\}
\]
for some $V'\subseteq V$ of size $\ell$. We show that $G-V'$ contains no
$k$-clique.

Suppose instead that $K\subseteq V\setminus V'$ is a clique of size $k$.
Let $A_K\subseteq A$ be the $\binom{k}{2}$ voters corresponding to the edges
of $K$, and define
\[
S=A_K\cup D,
\qquad
T=\{p_1\}\cup\{r_i\in P_3\mid v_i\in V\setminus K\}.
\]
Then $|S|=|V|-k=q+1$. Every voter in $D$ approves all projects in $T$.
Every edge represented in $A_K$ has both endpoints in $K$, so its
corresponding voter approves every $r_i$ with $v_i\notin K$. Thus, all voters
in $S$ commonly approve $T$. Moreover,
\[
\cost(T)
=\epsilon+(|V|-k)\frac{\epsilon}{q}
=2\epsilon+\frac{\epsilon}{q}.
\]
Because $K\cap V'=\emptyset$, voters in $A_K$ receive only $p_1,p_2$, while
voters in $D$ receive only $p_1,p_3$. Hence every voter in $S$ receives
exactly $2\epsilon$ from $P'$. Lemma~\ref{lem:ejr-violation}(5) now shows
that $(S,T)$ is an EJR violation, contradicting the assumption that $P'$
satisfies EJR. Therefore, $G-V'$ has no clique of size $k$.

The reduction uses $\alpha=\beta=1$ and applies to both
$BR\in\{\pairs,\cons\}$, proving $\Sigma_2^P$-hardness.

\section{Additional Experimental Details}
\subsection{Implementation and timing}\label{subsec:implementation}
The experiments were run in Python on a c6i.8xlarge EC2 instance for
approximately two hours, with up to 32 parallel workers. The two timing
tables compare specific implementations on the listed elections. Times
include preprocessing, model construction and solving or scoring, after
reading the election.

\paragraph{PAIRS.}
For an outcome \(W\), form the \(n\times |W|\) approval matrix \(M_W\).
Each entry of \(M_WM_W^\top\) counts the selected projects that two voters
both approve. Counting the positive entries strictly above the diagonal
gives PAIRS. The integer program instead uses a binary selection \(x_p\)
for each project and \(y_{uv}\) for each pair with a common approval. It
maximizes \(\sum_{\{u,v\}}y_{uv}\) subject to the budget and
\(y_{uv}\le\sum_{p\in A_u\cap A_v}x_p\). There are \(O(m+n^2)\) variables
and up to \(O(mn^2)\) nonzero coefficients. Table~\ref{tab:ilp-vs-enum}
reports the observed times.

\begin{table*}[htbp]
\centering
\caption{Elapsed seconds for PAIRS on five Toulouse elections in 2022.}
\label{tab:ilp-vs-enum}
\begin{threeparttable}
\begin{tabular}{lrrrrr}
\toprule
\textbf{Pabulib ID} & \textbf{Voters} & \textbf{Projects} & \textbf{Enum (s)} & \textbf{ILP (s)} & \textbf{Speedup} \\
\midrule
France\_Toulouse\_2022\_7  & 154 & 10 & 0.066 & 0.231 & 3.5$\times$ \\
France\_Toulouse\_2022\_13 & 304 & 10 & 0.188 & 0.397 & 2.1$\times$ \\
France\_Toulouse\_2022\_8  & 512 & 10 & 0.678 & 1.004 & 1.5$\times$ \\
France\_Toulouse\_2022\_12 & 659 & 10 & 0.623 & 2.777 & 4.5$\times$ \\
France\_Toulouse\_2022\_1  & 972 & 10 & 2.759 & 3.904 & 1.4$\times$ \\
\midrule
\textbf{Total} & --- & --- & \textbf{4.31} & \textbf{8.31} & \textbf{1.9$\times$} \\
\bottomrule
\end{tabular}
\begin{tablenotes}[para,flushleft]
\small
\item Speedup is the ratio of the unrounded running times.
\end{tablenotes}
\end{threeparttable}
\end{table*}

\paragraph{CONS.}
For each feasible subset, join each selected project's supporters in a
disjoint-set (union--find) structure. Summing \(\binom{|C|}{2}\) over the
resulting components \(C\) gives CONS. The comparison model uses the
potential voter edges \(E=\{\{u,v\}:A_u\cap A_v\ne\emptyset\}\).
A binary edge variable can be active only if the voters share a selected
project. For each pair of voters with nonempty ballots, a separate commodity
may send one unit of flow from one endpoint to the other. Every directed
arc is bounded by its undirected edge-activation variable, and flow is
conserved at the intermediate voters. The objective maximizes the number
of commodities routed.

For \(n_+\) nonempty ballots this encoding creates
\(2|E|\binom{n_+}{2}\) continuous flow variables, of order \(n^4\)
in a dense potential graph. In the two-project Marysin election in
Table~\ref{tab:ilp-vs-enum-cons}, 85 voters and 2{,}790 potential edges
give 19{,}920{,}600 flow variables. Enumeration scores four subsets.
The difference between 0.0002 and 390.3 seconds explains our choice on
these small elections; other formulations or larger project sets can have
different running times.

\begin{table*}[htbp]
\centering
\caption{Elapsed seconds for CONS on three elections.}
\label{tab:ilp-vs-enum-cons}
\begin{threeparttable}
\begin{tabular}{lrrrrr}
\toprule
\textbf{Pabulib ID} & \textbf{Voters} & \textbf{Projects} & \textbf{Enum (s)} & \textbf{ILP (s)} & \textbf{Speedup} \\
\midrule
poland\_warszawa\_2017\_plac & 27 & 4 & 0.0003 & 2.95 & 9{,}830$\times$ \\
poland\_warszawa\_2019\_marysin & 85 & 2 & 0.0002 & 390.3 & 1{,}839{,}144$\times$ \\
France\_Toulouse\_2022\_17 & 93 & 10 & 0.02 & 264.0 & 13{,}200$\times$ \\
\bottomrule
\end{tabular}
\begin{tablenotes}[para,flushleft]
\small
\item Speedup is the ratio of the unrounded running times.
\end{tablenotes}
\end{threeparttable}
\end{table*}

\subsection{Exact EJR separation and the empirical screen}\label{sec:ejr-test}
For a fixed outcome \(W\), let \(u_i=\cost(A_i\cap W)\).
For each nonempty project bundle \(T\) of cost at most \(b\), define
\[
 s_T(W)=|\{i\in N:T\subseteq A_i,\ u_i<\cost(T)\}|.
\]
A violation of \(\beta\)-EJR exists exactly when some \(T\) has
\(s_T(W)>0\) and \(n\cost(T)\le\beta b\,s_T(W)\). A violating coalition is
contained in this set of voters; taking all of them can only increase its
entitlement. Hence the supremum of the satisfied parameters equals
\[
 \min\left\{1,\ 
   \min_{T:\,s_T(W)>0}\frac{n\cost(T)}{b\,s_T(W)}\right\},
\]
with an empty inner minimum taken as \(+\infty\). At a binding ratio the
weak affordability inequality already yields a violation, so a supremum of
one alone does not certify ordinary EJR.

An exact separator for a proposed \(W,\beta\) can either enumerate \(T\) or
use binary variables \(s_i,z_p\) selecting a nonempty coalition and bundle.
Require \(s_i+z_p\le1\) whenever \(p\notin A_i\), and impose
\[
 \begin{aligned}
 n\sum_p c(p)z_p&\ \le\ \beta b\sum_i s_i,\\
 s_i=1&\ \Longrightarrow\ u_i<\sum_p c(p)z_p .
 \end{aligned}
\]
After scaling costs to integers, strict improvement requires at least one
unit. A feasible model returns a witness \((S,T)\). In a master problem
over outcomes, the constraint that at least one \(i\in S\) receive utility
at least \(\cost(T)\) eliminates this violation for that coalition and
bundle.

\paragraph{Empirical EJR screen.}
For a nonempty \(W\), the archived implementation tests thresholds \(q\)
drawn from the integer utilities plus one, individual project costs, prefix
sums of sorted project costs, and the budget, restricted to the range from
the minimum project cost to the budget. At each \(q\) it seeks the largest
group of voters with \(u_i<q\) who all approve projects of total cost
\emph{at least} \(q\). It uses \(nq/(b|S|)\) as the bound, takes the minimum
over returned groups, and clips the result at one. The code returns one for
the empty outcome or when no group is returned and can use an incumbent at
a solver time limit. We denote this recorded score by
\(\widehat\alpha_{\rm EJR}\). In the figures, passing the screen means
\(\widehat\alpha_{\rm EJR}\ge0.9999\).

For example, take a budget of 10, three projects costing 2, 6 and 7, and two
voters. The first approves the projects costing 2 and 6, the second approves
only the project costing 7. Fund the projects costing 2 and 7. Each voter
has a budget share of 5 and receives at least the cost of every individually
affordable approved bundle; they share no approvals. The outcome satisfies
EJR. Yet \(q=3\) is tested for the first voter, whose approved project of
cost 6 passes the screen's common-cost test. This gives a bound of
\(2\cdot3/10=0.6\). The threshold lies between indivisible bundle costs.
The EJR columns report the archived diagnostic; an exact empirical
classification would require applying the bundle test to the original ballots.

\paragraph{Rules and ties.}
MES uses greedy AV to spend any remaining budget after the equal-shares
phase. The greedy variants choose affordable projects by their respective
scores. The figures use one recorded outcome per rule and election; they
do not search over alternative tie resolutions for a rule. The attainable
row for AV and CC checks all feasible subsets. For each EJR-screen column,
it checks whether any bridging-maximizing outcome passes the recorded
screen.

\subsection{Additional empirical results}\label{subsec:additional-empirical-results}
\begin{figure}[tbp]
  \centering
  \includegraphics[width=\linewidth]{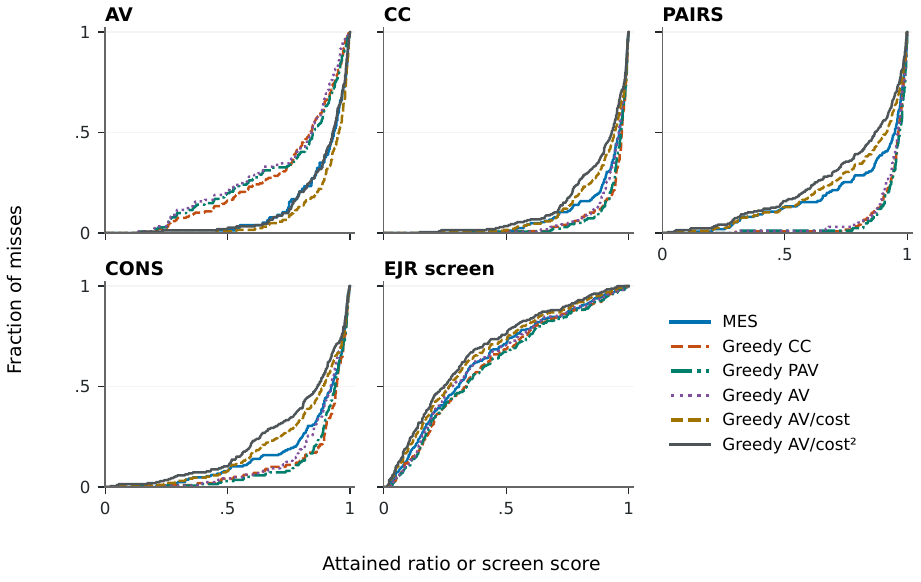}
  \caption{Empirical cumulative distributions conditional on missing the
  indicated target. At \(x\), the curve gives the fraction of misses with
  attained ratio (or EJR screen score) at most \(x\).}
  \label{fig:alpha_cdf}
\end{figure}

\paragraph{Outcomes below the maximum.}
Figure~\ref{fig:alpha_cdf} conditions on the elections where a rule misses a
particular objective. Thus a curve describes the size of its shortfalls,
and Figure~\ref{fig:metrics_bar} gives the complementary frequency of
exact hits. For the screen, the same 0.9999 cutoff is used in both figures.

\begin{figure}[tbp]
  \centering
  \includegraphics[width=\linewidth]{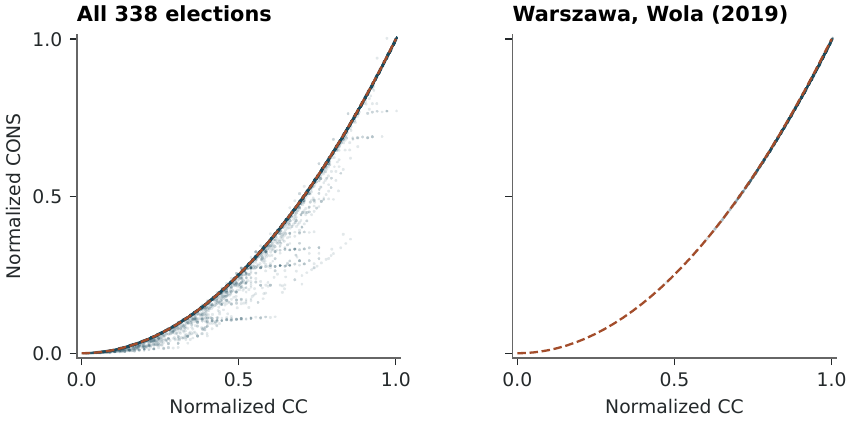}
  \caption{Normalized CONS against normalized CC for all nonempty outcomes
  (left) and for Warszawa, Wola--Rejon Po{\l}udniowy, 2019 (right).
  The line is \(y=x^2\).}
  \label{fig:cons_vs_cc_scatter}
\end{figure}

\paragraph{Connectivity and coverage.}
Across the 216{,}585 nonempty feasible outcomes, 215{,}276 (99.4\%) satisfy
\(|\alpha_{\cons}-\alpha_{\cc}^2|\le0.01\). In 310 of 338 elections
(91.7\%) every nonempty outcome meets this bound. Figure~\ref{fig:cons_vs_cc_scatter}
shows the pooled observations and one election in detail.

\begin{figure}[tbp]
  \centering
  \includegraphics[width=.84\linewidth]{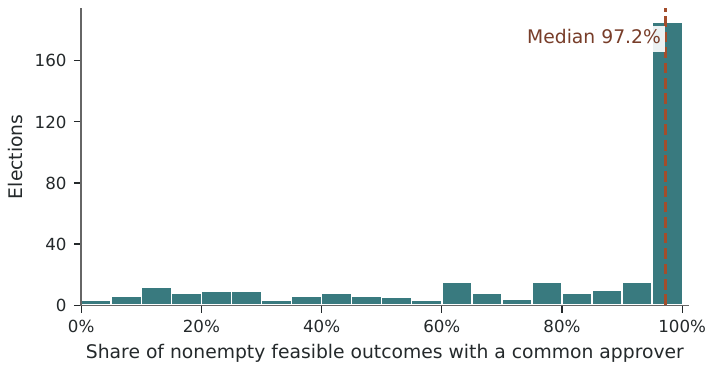}
  \caption{Distribution across the 338 elections of the proportion of
  nonempty feasible outcomes whose projects a single voter approves. Each
  election contributes one observation.}
  \label{fig:subset_superset_coverage}
\end{figure}

To count common approvers of a nonempty feasible \(W\), inclusion--exclusion
on the archived coverage scores gives
\(\sum_{\emptyset\ne T\subseteq W}(-1)^{|T|+1}\cc(T)\). Every subset used
in the sum is feasible. In 147 of 338 elections (43.5\%), every nonempty
feasible outcome has a common approver. The median within-election
proportion is 97.2\%; pooling outcomes instead gives 59.5\% because the
elections have different numbers of feasible outcomes
(Figure~\ref{fig:subset_superset_coverage}).

\FloatBarrier
\section{Challenges for Cost-Aware Bridging Objectives}
\label{app:cost-aware-bridging}

A natural way to incorporate project costs into PAIRS is to weight each
connected voter pair by the costs of the selected projects responsible for the
connection.  Two immediate variants are to use either the total cost or the
maximum cost of these projects.  For an outcome \(W\) and a voter pair
\(\{u,v\}\), let
\[
    C_{uv}(W):=A_u\cap A_v\cap W
\]
be the selected projects jointly approved by \(u\) and \(v\).  The two variants
can be written as
\[
    \pairs_{\mathrm{sum}}(W)
    :=
    \sum_{\{u,v\}\in N^{(2)}}
    \sum_{p\in C_{uv}(W)} c(p)
\]
and
\[
    \pairs_{\max}(W)
    :=
    \sum_{\substack{\{u,v\}\in N^{(2)}\\C_{uv}(W)\neq\emptyset}}
    \max_{p\in C_{uv}(W)} c(p).
\]
Both variants can behave counterintuitively.

\paragraph{Weighting by total cost.}
Changing the order of summation gives
\[
    \pairs_{\mathrm{sum}}(W)
    =
    \sum_{p\in W}c(p)\binom{|N_p|}{2}.
\]
Thus, the objective becomes an additive project score.  In particular, it does
not account for overlap between the voter pairs connected by different
projects: selecting multiple projects approved by the same group repeatedly
rewards the same pairs.  Under a budget constraint, the score per unit cost of
a project is simply \(\binom{|N_p|}{2}\), so the objective largely favors
projects with many approvals rather than projects that create new connections
between groups.

\paragraph{A common counterexample.}
Let the voters be partitioned into three groups \(X,Y,Z\), where
\(|X|=0.4n\) and \(|Y|=|Z|=0.3n\), and let the budget be \(B\).  Assume that
\(n=10k\) for some positive integer \(k\).  Consider four projects with the
following costs and supporter sets:
\[
\begin{array}{c|c|c}
    \text{project} & \text{cost} & \text{supporters}\\
    \hline
    p_1 & B   & X\cup Y\\
    p_2 & B/3 & X\cup Y\\
    p_3 & B/3 & Y\cup Z\\
    p_4 & B/3 & X\cup Z
\end{array}
\]
The outcomes \(W_1=\{p_1\}\) and \(W_2=\{p_2,p_3,p_4\}\) are both feasible.
The first connects only \(X\) and \(Y\), whereas the second directly connects
every pair of groups and, consequently, every pair of voters.

Under total-cost weighting,
\[
    \pairs_{\mathrm{sum}}(W_1)
    =
    B\binom{0.7n}{2},
\]
while
\[
    \pairs_{\mathrm{sum}}(W_2)
    =
    \frac{B}{3}
    \left(
        2\binom{0.7n}{2}+\binom{0.6n}{2}
    \right).
\]
Indeed, the difference between the former and the latter is
\(B(13k^2-k)/6>0\).  Thus, total-cost weighting prefers \(W_1\), despite
\(W_2\) connecting all three groups.

\paragraph{Weighting by maximum cost.}
The same example also causes difficulty for maximum-cost weighting.  Every
pair connected by \(W_1\) receives weight \(B\), so
\[
    \pairs_{\max}(W_1)=B\binom{0.7n}{2}.
\]
By contrast, \(W_2\) connects every voter pair, but each such pair receives
weight only \(B/3\).  Hence,
\[
    \pairs_{\max}(W_2)=\frac{B}{3}\binom{n}{2},
\]
and the difference is \(B(47k^2-11k)/6>0\).  The weighted objective therefore
again prefers the single project connecting two groups over the outcome that
connects all three groups.

These examples do not rule out cost-aware notions of bridging.  Rather, they
show that costs must be incorporated without rewarding redundancy or treating
project expense itself as evidence of a stronger connection.

\end{document}